\pdfoutput=1
\documentclass[authoryear,preprint,review,3p,12pt]{elsarticle}

\usepackage{amssymb}
\usepackage{amsmath}
\usepackage{subcaption}
\usepackage{hyperref}
\usepackage{float}
\usepackage{bm}
\usepackage{multirow}
\usepackage{makecell}
\usepackage{ragged2e}
\usepackage[authoryear]{natbib}  

\usepackage[utf8]{inputenc}
\usepackage{amsfonts}%
\usepackage{mathrsfs}%
\usepackage[title]{appendix}%
\usepackage{xcolor}%
\usepackage{textcomp}%
\usepackage{manyfoot}%
\usepackage{booktabs}%
\usepackage{algorithm}%
\usepackage{algorithmicx}%
\usepackage{algpseudocode}%
\usepackage{listings}%
\DeclareCaptionLabelFormat{new}{\textup{(}{\text{#2}}\textup{)}}
\usepackage{overpic}

\begin{document}

\begin{frontmatter}

\title{A nonlinear transient finite element analysis of coupled solvent diffusion and large deformation of magneto-elastic hydrogel balloon}

\author[inst1]{Shaikh Nadeem Karim}

\affiliation[inst1]{organization={Department of Applied Mechanics and Biomedical Engineering},%
            addressline={Indian Institute of Technology Madras}, 
            city={Chennai},
            postcode={600036}, 
            country={India}}

\author[inst1]{Ganesh Tamadapu}

\begin{abstract}

This study investigates the transient swelling of an initially spherical magnetoelastic
hydrogel balloon. The balloon is inflated at constant internal pressure and exposed to the non-uniform magnetic field of one or two current-carrying coils. The gel is modelled as a
compressible Ogden network mixed with solvent through Flory--Huggins theory, with an
additive elastic--mixing--magnetic decomposition of the free energy. A fully coupled nonlinear finite element formulation uses a staggered solution scheme and a Taylor–Hood-type discretisation. It follows the balloon from its dry inflated state to either swollen equilibrium or burst. The formulation recovers the analytical free-swelling stretch and the delayed-burst regime, and its converged states agree with equilibrium energy minimisation. Symmetry-breaking (pear-shaped) dry configurations do not survive swelling: they rearrange rapidly and then relax by diffusion onto the symmetric branch. Transient membrane stresses can exceed their equilibrium values by nearly an order of magnitude. For coils which are sufficiently close to undeformed balloon, fields beyond a critical intensity close the slow-inflation curve into a loop. Stable swollen equilibria then exist only within a bounded pressure window that narrows as the field increases, and outside of it the balloon bursts. Under two coaxial coils, asymmetry about the equator appears only when the coils differ in current or radius. The magnetic field therefore acts as a control parameter for shape of balloon and permits or prevents stable swelling at a given pressure. 
\end{abstract}


\end{frontmatter}

\section{Introduction} 
Hydrogels are polymeric materials with a crosslinked network that can absorb large amounts of fluid without disrupting the skeletal structure of the network. Together, the polymeric network and the absorbed fluid form a swollen aggregate known as a polymeric gel; when the fluid is water, the gel is called a hydrogel. This composition gives hydrogels a rare combination of large deformability, fluid permeability, and biocompatibility, and is also what makes them natural models for soft biological tissues, which typically contain $60$--$90\%$ water and may themselves be regarded as biological hydrogels \citep{liu2020hydrogel}. The degree of swelling and the resulting deformation can be controlled by a variety of external stimuli, including mechanical loading \citep{chester2010coupled}, humidity \citep{lv2018humidity}, temperature \citep{chester2011thermo,yoon2010comparison}, and pH \citep{reyes2014ph}. These constitutive features, combined with the geometric nonlinearity inherent to highly deformable, inflatable thin-walled structures, make the inflation and instability of hydrogel thin shells such as hydrogel balloons a particularly rich problem in nonlinear elasticity.

Thin-walled structures made of permeable gel-like materials in an aqueous environment are ubiquitous in nature. Biological cells are perhaps the best-known example \citep{jung2020nonlinear,mills2004nonlinear}, but many other tissues can also be viewed as hydrogel shells. The zona pellucida is a hydrogel glycoprotein coating that envelopes mammalian oocytes and early embryos \citep{priel2024hyperelastic}; the gastric mucus layer is a mucin-based gel barrier that lines the stomach \citep{phillipson2008gastric}; the airway surface liquid is a layered hydrogel film that coats the lung epithelium \citep{rosenzweig2002capillary}; and in the eye, the Bowman layer of the cornea and lens capsule are thin permeable hydrogel shells under sustained mechanical load \citep{hollman2002strain,fisher1969elastic}.  Engineered hydrogel thin shells, especially hydrogel capsules, have also found a growing range of biomedical uses, including cell encapsulation \citep{orive2003cell,lanza1996encapsulated}, the study of tumor growth under mechanical confinement \citep{alessandri2013cellular}, and controlled drug delivery \citep{luo2014near}.

The inflation of hydrogel thin-walled structures is therefore of interest on three fronts. From a nonlinear-elasticity standpoint, it brings together geometric nonlinearity, material nonlinearity, and coupled diffusion--deformation physics in a single problem. From a biology standpoint, many of the soft, fluid-permeable shells found in living systems can be modeled within the same framework. From a biomedical-engineering standpoint, the same analysis underpins the design and behavior of hydrogel capsules and related shell-based devices.

However, a practical limitation of conventional hydrogel shells is that the stimuli driving their response, mechanical loading, humidity, temperature, and pH, are largely passive and that the resulting deformation is governed by solvent diffusion, which is slow. \emph{Magneto-active hydrogels} (MAHs), formed by embedding magnetic nanoparticles within the crosslinked polymer network \citep{manish2022synthesis,manish2023synthesis}, address both limitations: they deform on demand in response to an externally applied magnetic field, and the resulting magneto-mechanical actuation can be controlled remotely, locally, and on time scales much shorter than the diffusive ones intrinsic to the gel. The price of this added controllability is that the magneto-mechanical response is strongly coupled to solvent transport, so any predictive model must treat the magnetic field, the elastic deformation, and the solvent diffusion together.

The theoretical foundations for the coupled diffusion--deformation response of hydrogels were laid well before magnetic actuation entered the picture. \citet{gibbs1878pages} formulated an equilibrium thermodynamic theory of an elastic solid that absorbs a fluid, and \citet{biot1941general} introduced the linear poroelastic framework, combining elasticity with Darcy's law, which underlies modern soil-consolidation analyzes. For polymer--solvent systems specifically, \citet{flory1943statistical} derived the mixing free energy of long crosslinked chains with small-molecule solvent: as the solvent enters the network, the configurational entropy of the chains decreases while that of the mixture increases, and equilibrium between the gel and the surrounding bath follows from a balance of the two. Building on this picture, \citet{hong2008theory} developed a nonequilibrium finite-deformation theory of coupled diffusion and large deformation in polymer gels, which has since become the standard reference. A variational saddle-point formulation of this theory is given in \citet{boger2017minimization}, and a phase-field extension to fracture in hydrogels has been developed in the same framework \citep{boger2017minimization}. \citet{chester2010coupled} proposed an alternative finite-deformation theory in which the elastic stain energy is taken from a non-Gaussian statistical-mechanical model that captures the limited extensibility of polymer chains; \citet{chester2011thermo} subsequently generalized it to thermally responsive elastomeric gels.

For thin-walled hydrogel structures specifically, \citet{zhang2011instability} studied the instabilities of hydrogel plates using the finite-element method, while \citet{TengLi_JMPS2019} and \citet{TAMADAPU2022} analyzed the delayed burst of spherical and toroidal hydrogel balloons, respectively, identifying the diffusion-controlled regime in which a balloon held at constant pressure slowly inflates until catastrophic failure.

On the magneto-mechanical side, \citet{garcia2020magneto} formulated a general continuum framework for the coupled magneto--diffusion--deformation response of hydrogels and implemented it within a finite-element solver (FEniCS) for three-dimensional bulk problems. More recently, \citet{gonzalez2024magneto} fabricated an ultra-soft magnetoelastic hydrogel from human blood plasma and quantified the effect of magnetic actuation on solvent expulsion and intra-network diffusion dynamics. The literature on bulk magnetoelastic hydrogels is by now reasonably mature, with several constitutive frameworks and three-dimensional boundary-value problems solved within nonlinear multiphysics finite-element settings.

What remains conspicuously absent is the corresponding analysis for \emph{thin-walled} magnetoelastic hydrogel structures. To the best of our knowledge, there is no transient nonlinear finite-element study of the inflation of a pressurized magnetoelastic hydrogel balloon under an external magnetic field, and no systematic account of how diffusion and magnetic field intensity together control the time-dependent swelling, deformation, and instability of such a structure. The present work fills this gap. We develop a transient, fully coupled, nonlinear multiphysics finite-element formulation for the inflation of a spherical magnetoelastic hydrogel balloon held at constant pressure and subjected to the non-uniform magnetic field of a current-carrying coil, and we use it to characterize how the magnetic field intensity and the diffusion dynamics jointly shape the balloon's evolution from its dry inflated state to its swollen equilibrium.

The manuscript is organized as follows. Section~\ref{sec: Kinematics of deformation} presents the kinematics of the axisymmetric deformation of a compressible magnetoelastic spherical hydrogel balloon subjected to a non-uniform magnetic field. Section~\ref{sec:Total potential Energy} gives the total potential energy of the inflation problem of a spherical magnetoelastic gel balloon.  Section~\ref{Method_II} develops the weak forms of the linear momentum balance and fluid transport for the coupled diffusion--deformation-magnetic problem and presents a staggered, transient, nonlinear finite element algorithm for balloon inflation under applied pressure and magnetic loading. Section~\ref{sec: result} discusses the results. Section~\ref{sec: absence of magnetic field} examines balloon swelling in the absence of a magnetic field: section~\ref{sec: free_swelling} addresses free swelling, Section~\ref{sec: Swelling of pressurized balloons} considers the swelling of pressurized balloons, and section~\ref{sec: Delayed_burst} describes the delayed-burst phenomenon. Section~\ref{sec: Magnetic} presents results under the influence of a magnetic field: section~\ref{sec:Symmetric magnetic loading} shows the swelling of pressurized, symmetry-preserving balloons, while section~\ref{sec: Symmetry-breaking magnetic time evolution} explains the swelling of symmetry-breaking balloons under a magnetic field. Section~\ref{sec:two_coil} presents results for swelling of gel balloon under the application of magnetic field created by two coils, where symmetric as well as asymmetric magnetic loading is considered. Section~\ref{sec:Appendix} formulates the total potential energy, accounting for the magnetic, pressure, strain, and mixing energy contributions; it then derives the governing differential equations from the principle of minimum potential energy and describes the procedure for solving the resulting system of ODEs to obtain the equilibrium solution for the final equilibrium configuration, which marks the end of the diffusion process.
\section{Kinematics of axisymmetric deformation} 
\label{sec: Kinematics of deformation} 
\begin{figure}[h] 
    \centering
    \includegraphics[width=\linewidth]{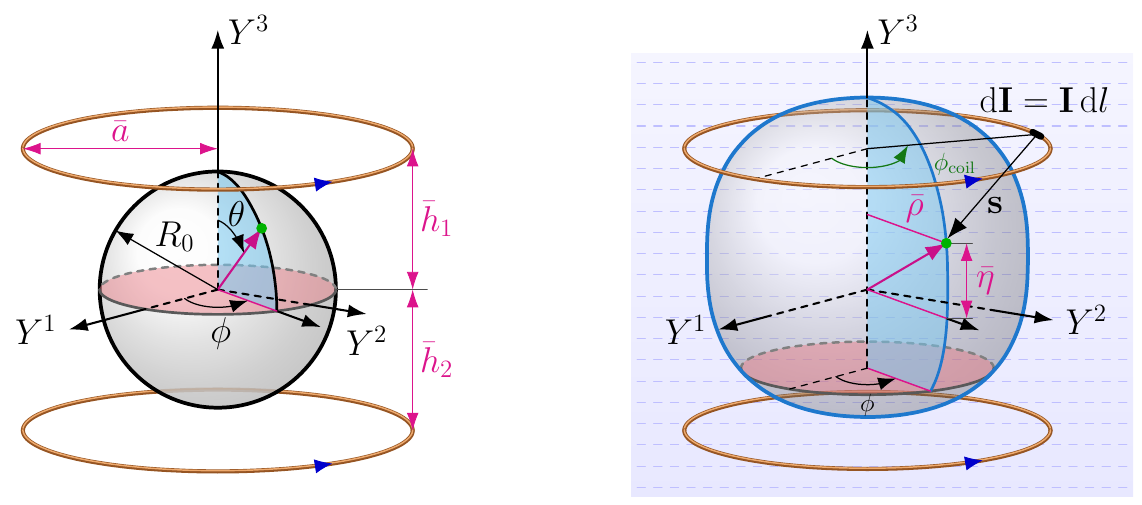}
    \caption{Initially dry spherical balloon (left) and its swollen, axisymmetrically deformed configuration (right) when immersed in a solvent bath and subjected to internal pressure and the magnetic field produced by the current-carrying coil(s).}
    \label{fig:enter-label1}
\end{figure}
In this work, we study the finite inflation of a spherical balloon
made of a compressible magnetoelastic gel. In the
undeformed dry configuration, the balloon has mid-surface radius
$R_0$ and uniform thickness $H_0$. The deformed configuration is
assumed to be axisymmetric about the $Y^3-$ axis
(Fig.~\ref{fig:enter-label1}). The problem is formulated in a
spherical coordinate system $(\theta, \varphi, \zeta)$, in which
$\theta$ and $\varphi$ are the polar and azimuthal angles on the
mid-surface and $\zeta$ is the through-thickness coordinate measured
from the mid-surface. Let $\mathbf{X}$ be the position vector of a material
point of the gel on the mid-surface ($\zeta=0$) in the undeformed dry
configuration. Let $\mathbf{x}$ be the position vector of the same
point in a continuously swelling current configuration. The expressions for $\mathbf{X}$ and $\mathbf{x}$ are given by   
\begin{align}
\begin{split}
  \mathbf{X}(\theta,\varphi)&= R_0 \sin{\theta} \cos{\varphi} \, \mathbf{E}_1 + R_0 \sin{\theta} \sin{\varphi} \, \mathbf{E}_2 + R_0 \cos{\theta} \, \mathbf{E}_3, \\ 
  {\mathbf{x}(\theta,\varphi,t)}& = {\bar{\rho}\left(\theta,t \right) \cos{\varphi} \, \mathbf{e}_1 + \bar{\rho}\left(\theta ,t\right) \sin{\varphi} \, \mathbf{e}_2 + \bar{\eta}\left(\theta ,t\right) \, \mathbf{e}_3,}   \label{Position vec} 
\end{split}   
\end{align}
where $\bar{\rho}(\theta,t)$ and $\bar{\eta}(\theta,t)$ are,
respectively, the horizontal and vertical components of $\mathbf{x}$ in a
meridional plane (a plane of constant $\varphi$), as shown in
Fig.~\ref{fig:enter-label1}. The orthonormal triads
$\{\mathbf{E}_i\}$ and $\{\mathbf{e}_i\}$ are the bases of the
Euclidean coordinate systems $Y^i$ and $y^i$ associated with the
undeformed (reference) and deformed (current) configurations,
respectively. The deformation mapping $\mathcal{X}$ between the two
configurations is defined by $\mathbf{x}(\theta,\varphi,t)=\mathcal{X}\!\bigl(\mathbf{X}(\theta,\varphi),\, t\bigr).$
The bases vectors \textcolor{black}{$\mathbf{G}_{\alpha}$} and \textcolor{black}{$\mathbf{g}_{\alpha}$, $\alpha=\{1,2\},$} for the undeformed and deformed mid-surface, respectively, are given as, 

\begin{equation}
   { \mathbf{G}_{\alpha} = \frac{\partial \mathbf{X}}{\partial q^{\alpha}},} \quad {\mathbf{g}_{\alpha} = \frac{\partial \mathbf{x}}{\partial q^{\alpha}}}  \quad \text{where} \, \left(q^1,q^2\right) = \left(\theta , \varphi \right).  
    \label{tangentC vec} 
\end{equation} 
The surface normal vector $\mathbf{N}$ in the undeformed configuration at point $\mathbf{x}$ and the surface normal vector $\mathbf{n}$ in the deformed configuration at point $\mathbf{x}$ are given, respectively, as,  

\begin{equation}
\mathbf{N} = \frac{ \mathbf{G}_1 \times \mathbf{G}_2}{|\mathbf{G}_1 \times \mathbf{G}_2 |} \quad \text{and} \quad \mathbf{n} = \frac{ \mathbf{g}_1 \times \mathbf{g}_2}{|\mathbf{g}_1 \times \mathbf{g}_2 |}. 
\label{current normal vec}
\end{equation}
The vector triads ($\mathbf{G}_1$, $\mathbf{G}_2$, $\mathbf{N}$) and ($\mathbf{g}_1$, $\mathbf{g}_2$, $\mathbf{n}$) form the bases in the curvilinear coordinate system ($\theta$, $\varphi$, $\zeta$) for the undeformed and deformed configurations, respectively. The quantities \textcolor{black}{$G_{\alpha \beta} = \mathbf{G}_\alpha \cdot \mathbf{G}_\beta$} and \textcolor{black}{$g_{\alpha\beta} = \mathbf{g}_\alpha \cdot \mathbf{g}_\beta$} for \textcolor{black}{$\alpha,\beta=\{1,2\}$} represent the covariant metric tensor components for the undeformed and deformed configurations, respectively. The metric tensors for the mid--surface, $\mathbf{G}$ (undeformed) and $\mathbf{g}$ (deformed) are expressed as,  
\begin{equation}
\textcolor{black}{\mathbf{G} = R_0^2\,\mathbf{G}^1 \otimes \mathbf{G}^1 +  R_0^2 \sin^2\theta \,\mathbf{G}^2 \otimes \mathbf{G}^2,}  
\label{Gmetric} 
\end{equation} 
\begin{equation}
\textcolor{black}{\mathbf{g} = \left(\bar{\rho}_{,\theta}^2 + \bar{\eta}_{,\theta}^2 \right)\, \mathbf{g}^1 \otimes \mathbf{g}^1 +  \bar{\rho}^2\,\mathbf{g}^2 \otimes \mathbf{g}^2.} 
\label{gmetric} 
\end{equation} 
The components of the inverse of the metric tensors $\mathbf{G}$ and $\mathbf{g}$ are given as  $G^{-1}_{\alpha \beta} = G^{\alpha \beta}$, and $g^{-1}_{\alpha \beta}  = g^{\alpha \beta}$, respectively. The contravariant basis vectors in the undeformed and deformed configurations are defined, respectively, as $\mathbf{G}^\alpha =G^{\alpha\beta}\mathbf{G}_\beta, \mathbf{g}^\alpha = g^{\alpha\beta} \mathbf{g}_\beta.$
\textcolor{black}{The deformation gradient tensor $\mathbf{F}$, which maps the undeformed mid-surface configuration to the deformed mid-surface, is given by $\mathbf{F} = \mathbf{g}_{\beta} \otimes \mathbf{G}^{\beta}$. The left Cauchy-Green tensor $\mathbf{C}$, associated with the mid-surface, is expressed as $\mathbf{C} = \mathbf{F}^T \mathbf{F} = g_{\alpha \beta} \mathbf{G^\alpha} \otimes \mathbf{G^\beta} = g_{\alpha \beta} G^{\alpha \gamma} \mathbf{G}_{\gamma} \otimes \mathbf{G}^\beta$. The Cauchy-Green tensor for the mid--surface of the balloon can be expressed in terms of mixed basis as:} 
\begin{align}
\mathbf{C} =\frac{\bar{\rho}_{,\theta}^2 + \bar{\eta}_{,\theta}^2 }{R_0^2}\, \mathbf{G}_1\otimes\mathbf{G}^1 + \frac{\bar{\rho}^2}{R_0^2 \sin^2\theta}\, \mathbf{G}_2\otimes\mathbf{G}^2. 
\end{align}
Let the eigenvalues of the Cauchy--Green tensor $\mathbf{C}$ be $\lambda_1^2$ and $\lambda_2^2$, where $\lambda_1$ and $\lambda_2$ are the in-plane principal stretch ratios in $\theta$ and $\phi$ directions, respectively. The expressions for the stretch ratios are: 
\begin{equation}
    \lambda_1 = \frac{ \sqrt{\bar{\rho}_{,\theta}^2 + \bar{\eta}_{,\theta}^2} }{R_0}, \quad \lambda_2 = \frac{\bar{\rho}}{R_0 \sin\theta}.    
\end{equation} 
Consider the gel balloon membrane as a three-dimensional bulk
continuum, and let $\mathbf{C}_{\text{bulk}}$ denote the left
Cauchy--Green tensor of the bulk deformation. Because the gel is
modelled as a mixture of an incompressible polymer network and an
incompressible solvent, any change in volume during
deformation must be due solely to a change in the local solvent
content. Following the argument used for a compressible gel in
\citep{chester2010coupled, chester2011thermo}, we therefore obtain
the swelling constraint $J=\det\mathbf{F}=1 + \bar\nu\,\bar C_r,$
where $\bar\nu$ is the molar volume of the solvent and $\bar C_r$ is the
number of moles of solvent absorbed per unit reference volume of
the elastomer. Equivalently, in terms of the principal components
$(\lambda_1^2,\lambda_2^2,\lambda_3^2)$ of $\mathbf{C}_{\text{bulk}}$,
$\det\mathbf{C}_{\text{bulk}}=(\lambda_1\lambda_2\lambda_3)^{2}=(1 + \bar\nu\,\bar C_r)^{2}.$
Hence, principal stretch in the direction normal to
the mid-surface, $\lambda_3$, given as, 
\begin{equation*}
    \lambda_3 \;=\; \frac{1 + \bar\nu\,\bar C_r}{\lambda_1\,\lambda_2}.
\end{equation*} 

\section{Total potential Energy} 
\label{sec:Total potential Energy}
The total potential energy functional governing the axisymmetric
inflation of the magnetoelastic gel balloon is
\begin{align}
    \bar{\pi}(\mathbf{C}, \bar{C}_r, \mathbf{H})
    \;=\; \int_{A_0} \bar{S}\, R_0^{2}\sin\theta\,\bar{H}_0\,
            \mathrm{d}\theta\,\mathrm{d}\phi
    \;-\; \int_{A_0} \frac{\bar{P}}{2}\,\bar{\rho}^{2}\,\bar{\eta}_{,\theta}\,
            \mathrm{d}\theta\,\mathrm{d}\phi,
    \label{Total_potential_1}        
\end{align}
where the first term is the strain-energy contribution
(with $\bar{S}$ the free-energy density per unit reference volume and
$\bar{H}_0$ the mid-surface thickness) and the second term is the work
done by the internal pressure $\bar{P}$. Following
\citep{garcia2020magneto}, the free-energy density admits the additive
decomposition
\begin{equation}
    \bar{S} \;=\; \bar{S}_e + \bar{S}_{\mathrm{mix}} + \bar{S}_{\mathrm{mag}},
    \label{mag_gel_energy}
\end{equation}
where $\bar{S}_e$, $\bar{S}_{\mathrm{mix}}$, and $\bar{S}_{\mathrm{mag}}$ are the polymer-network, polymer-solvent mixing, and magnetic
contributions, respectively.

The polymer network is modeled as a compressible Ogden material,
\begin{align}
    \bar{S}_e
    = \sum_{i=1}^{3}\frac{\bar{\mu}_i}{\alpha_i}
        \!\left(\lambda_1^{\alpha_i} + \lambda_2^{\alpha_i}
                + \lambda_3^{\alpha_i} - 3\right)
    + \frac{\bar{\kappa}}{\beta^{2}}\!
        \left[\beta\log(1+\bar{C}_r\bar{\nu})
              + (1+\bar{C}_r\bar{\nu})^{-\beta} - 1\right],
    \label{energy_Ogden1}
\end{align}
where $\bar{\mu}_i$ are material constants, $\alpha_i$ are shape parameters, and $\bar{\kappa}$ and $\beta$ govern the volumetric
response. The Ogden parameters are related to the small-strain shear
modulus $\mu$ by $2\mu = \sum_{i=1}^{3}\bar{\mu}_i\alpha_i$, and the energy is non-dimensionalised by $\mu^{*}=\mu/2$.

The polymer-solvent mixing energy is taken from Flory--Huggins theory
\citep{flory1942thermodynamics, huggins1941solutions},
\begin{align}
    \bar{S}_{\mathrm{mix}}
    = R T\, \bar{C}_r\!\left[\frac{\chi}{\bar{C}_r\bar{\nu}+1}
        + \log\!\left(\frac{\bar{C}_r\bar{\nu}}{\bar{C}_r\bar{\nu}+1}\right)\right],
\end{align}
in which $R$ is the universal gas constant, $T$ the absolute
temperature, and $\chi$ the Flory--Huggins interaction parameter.

The magnetic energy density is taken to be
\begin{equation}
    \bar{S}_{\mathrm{mag}}
    \;=\; \frac{1+\bar{C}_r\bar{\nu}}{2}\,\mu_0\,\chi_m\,
          \mathbf{h}_a\!\cdot\!\mathbf{h}_a
    \;=\; \frac{1+\bar{C}_r\bar{\nu}}{2}\,\mu_0\,\chi_m\,
          \mathbf{C}^{-1}\mathbf{H}_a\!\cdot\!\mathbf{H}_a,
    \label{mag energy}
\end{equation}
where $\mu_0$ is the vacuum permeability, $\chi_m$ the magnetic
susceptibility. In \eqref{mag energy}, $\mathbf{h}_a$ and $\mathbf{H}_a = \mathbf{F}^\text{T}\mathbf{h}_a$ are the applied
magnetic field in the deformed and reference configurations,
respectively. 

Following \citep{shaikh2025inflation, PsuxenaIJNM2017},
the field $\mathbf{h}_a$ produced by a circular current-carrying coil of
radius $\bar{a}$ at height $\bar{h}$ above the equatorial plane is
\begin{align}
    \mathbf{h}_a
    = \frac{I}{4\pi}\int_0^{2\pi}
    \frac{(\bar{\eta}-\bar{h})\,\bar{a}\cos\varphi_{\mathrm{coil}}\,\mathbf{e}_1
        + (\bar{\eta}-\bar{h})\,\bar{a}\sin\varphi_{\mathrm{coil}}\,\mathbf{e}_2
        + \bigl(\bar{a}^{2}-\bar{\rho}\bar{a}\cos(\varphi_{\mathrm{coil}}-\varphi)\bigr)\,\mathbf{e}_3}
    {\bigl(\bar{\rho}^{2}+(\bar{\eta}-\bar{h})^{2}+\bar{a}^{2}-2\bar{\rho}\bar{a}\cos(\varphi_{\mathrm{coil}}-\varphi)\bigr)^{3/2}}
    \,\mathrm{d}\varphi_{\mathrm{coil}}.
    \label{ha calc}
\end{align}
Introducing the non-dimensional quantities 
$(\rho,\,\eta,\,a,\,h) = (\bar{\rho},\,\bar{\eta},\,\bar{a},\,\bar{h})/R_0$,
the applied field \eqref{ha calc} can be rewritten as
\begin{align}
    \mathbf{h}_a
    = \frac{I}{4\pi R_0}\int_0^{2\pi}
    \frac{(\eta-h)\,a\cos\varphi_{\mathrm{coil}}\,\mathbf{e}_1
        + (\eta-h)\,a\sin\varphi_{\mathrm{coil}}\,\mathbf{e}_2
        + \bigl(a^{2}-\rho a\cos(\varphi_{\mathrm{coil}}-\varphi)\bigr)\,\mathbf{e}_3}
    {\bigl(\rho^{2}+(\eta-h)^{2}+a^{2}-2\rho a\cos(\varphi_{\mathrm{coil}}-\varphi)\bigr)^{3/2}}
    \,\mathrm{d}\varphi_{\mathrm{coil}}.
    \label{ha calcnd}
\end{align}
Introducing the dimensionless magnetic-loading parameter $\mathcal{M}$ and
the dimensionless field magnitude $\mathcal{H}$ defined as
\begin{align}
    &\mathcal{M} \;=\; \frac{\mu_0\,I^{2}}{16\pi^{2}\,R_0^{2}\,\mu^{*}},
    \qquad \\  \ \nonumber
    &\mathcal{H}
    \;=\; \left|\int_0^{2\pi}\!
    \frac{(\eta-h)\,a\cos\varphi_{\mathrm{coil}}\,\mathbf{e}_1
        + (\eta-h)\,a\sin\varphi_{\mathrm{coil}}\,\mathbf{e}_2
        + \bigl(a^{2}-\rho a\cos(\varphi_{\mathrm{coil}}-\varphi)\bigr)\,\mathbf{e}_3}
    {\bigl(\rho^{2}+(\eta-h)^{2}+a^{2}-2\rho a\cos(\varphi_{\mathrm{coil}}-\varphi)\bigr)^{3/2}}
    \,\mathrm{d}\varphi_{\mathrm{coil}}\right|,
    \label{H cali}
\end{align} 
the magnetic part of the strain energy can be written as 

\begin{equation}
    \label{Sm_redx}
    \bar{S}_{\text{m}} =  \frac{J}{2} \chi_m \mu^* \mathcal{M} \mathcal{H}^2  = \frac{1}{2\phi} \chi_m \mu^* \mathcal{M} \mathcal{H}^2. 
\end{equation} 
Using the swelling constraint $1+\bar{C}_r\bar{\nu}=J=\phi^{-1}$,
where $\phi$ is the polymer volume fraction, the free-energy density
$\bar{S}$ can equivalently be written as $\bar{S}(\mathbf{C},\mathbf{H},J)$
or $\bar{S}(\mathbf{C},\mathbf{H},\phi)$. In terms of $J$,
\begin{align}
\begin{split}
    \bar{S}(\mathbf{C},\mathbf{H},J)
    \;=\; & \sum_{i=1}^{3}\frac{\bar{\mu}_i}{\alpha_i}\!
        \left[\lambda_1^{\alpha_i} + \lambda_2^{\alpha_i}
              + \!\left(\frac{J}{\lambda_1\lambda_2}\right)^{\!\alpha_i}
              - 3\right]
    + \frac{J}{2}\,\mu_0\,\chi_m\,
        \mathbf{C}^{-1}\mathbf{H}_a\!\cdot\!\mathbf{H}_a \\
    & + \frac{\bar{\kappa}}{\beta^{2}}\!
        \left[\beta\log J + J^{-\beta} - 1\right]
    + \frac{R T}{\bar{\nu}}\!
        \left[\frac{\chi}{J} + \log\!\frac{J-1}{J}\right](J-1),
\end{split}
\end{align}
and in terms of $\phi$,
\begin{align}
\begin{split}
    \bar{S}(\mathbf{C},\mathbf{H},\phi)
    \;=\; & \sum_{i=1}^{3}\frac{\bar{\mu}_i}{\alpha_i}\!
        \left[\lambda_1^{\alpha_i} + \lambda_2^{\alpha_i}
              + \!\left(\frac{1}{\phi\lambda_1\lambda_2}\right)^{\!\alpha_i}
              - 3\right]
    + \frac{\mu_0\,\chi_m}{2\,\phi}\,
        \mathbf{C}^{-1}\mathbf{H}_a\!\cdot\!\mathbf{H}_a \\
    & + \frac{\bar{\kappa}}{\beta^{2}}\!
        \left[\beta\log\!\frac{1}{\phi} + \phi^{\beta} - 1\right]
    + \frac{R T}{\bar{\nu}}\!
        \left[\phi\,\chi + \log(1-\phi)\right]\!
        \left(\frac{1}{\phi} - 1\right).
\end{split}
\end{align}
For brevity, the explicit $\theta$-dependence of $\lambda_1$,
$\lambda_2$, $J$, $\phi$, and $\mathbf{H}_a$ has been suppressed.

The total potential energy $\bar{\pi}$ from \eqref{Total_potential_1}  can be written as, 
\begin{equation}
    \bar{\pi} = \int_0^{2\pi} \int_0^{\pi} \left[ \left( \bar{S}_{\text{e}} - \bar{S}_{\text{m}} - \bar{S}_{\text{mix}} \right)R_0^{2}\sin\theta\,\bar{H}_0  + \frac{1}{2} \bar{P}\bar{\rho}^2\bar{\eta}_{,\theta}\right]   {\rm d}\theta \,{\rm d}\phi, 
    \label{E_T}
\end{equation}
Substituting  $\bar{S}_{\text{m}}$ from  \eqref{Sm_redx} in \eqref{E_T},   

\begin{equation}
\label{var energy1} 
    \bar{\pi} = \int_0^{2\pi} \int_0^{\pi} \left[ \left( {\bar{S}}_{\text{e}} - \bar{S}_{\text{mix}} -\frac{J}{2}  \chi \mu^* \mathcal{M} \mathcal{H}^2 \right)\bar{H}_0R_0^2 \sin \theta  + \frac{1}{2}\bar{P}{\bar{\rho}}^2{\bar{\eta}}_{,\theta}\right]\,{\rm d}\theta \,{\rm d}\phi,  
\end{equation}
 $\bar{\pi}$ is a functional in terms of primary field variables and independent variable $\theta$ i.e. $\bar{\pi} = \bar{\pi}\left(\bar{\rho}, \bar{\rho}_{,\theta}, \bar{\eta}, \bar{\eta}_{,\theta}, J,\theta \right)$.

\section{Finite element formulation}
\label{Method_II}
\subsection{Balance of linear momentum}

The strong form of the equilibrium equation at every point in the reference volume $\Omega$ of the balloon membrane reads
\begin{equation}
    \mathrm{Div}\,\mathbf{T} + \mathbf{b} = \mathbf{0},
    \label{eq:strong_form}
\end{equation}
where $\mathbf{T}$ is the first Piola--Kirchhoff stress and $\mathbf{b}$ is the body force per unit reference volume, supplemented by the traction boundary condition
\begin{equation}
    \mathbf{T}\,\mathbf{N} = \mathbf{t} \quad \text{on } \partial\Omega,
    \label{eq:traction_bc}
\end{equation}
in which $\mathbf{N}$ denotes the outward unit normal in the reference configuration.

To derive the weak form, we take the inner product of \eqref{eq:strong_form} with a kinematically admissible virtual displacement $\delta\mathbf{u}$ over $\Omega$:
$$ \int_\Omega \bigl(\mathrm{Div}\,\mathbf{T} + \mathbf{b}\bigr)\cdot\delta\mathbf{u}\,\mathrm{d}V = 0.$$
Applying the product identity $\mathrm{Div}\,\mathbf{T}\cdot\delta\mathbf{u} = \mathrm{Div}\bigl(\mathbf{T}^{\mathsf T}\delta\mathbf{u}\bigr) - \mathbf{T}:\nabla\delta\mathbf{u}$ and invoking the divergence theorem and substituting the traction boundary condition \eqref{eq:traction_bc} yields the principle of virtual work,
\begin{equation}
    \int_\Omega \mathbf{T}:\nabla\delta\mathbf{u}\,\mathrm{d}V \;=\; \int_{\partial\Omega} \mathbf{t}\cdot\delta\mathbf{u}\,\mathrm{d}A \;+\; \int_\Omega \mathbf{b}\cdot\delta\mathbf{u}\,\mathrm{d}V.
    \label{eq:virtual_work}
\end{equation}
In \eqref{eq:virtual_work}, the left-hand side represents the internal virtual work and the right-hand side represents the external virtual work.

For a hyperelastic material, $\mathbf{T}:\delta\mathbf{F} = \delta U$, where $U$ is the strain-energy density per unit reference volume. Integrating through the membrane thickness $\bar{H}_0$ replaces the volume integral on the left-hand side of \eqref{eq:virtual_work} with a surface integral over the reference mid-surface. Neglecting body forces and noting that the only external traction is the inflation pressure $\bar{P}$, which acts along the inward normal $-\mathbf{n}$ of the current surface, the virtual-work statement reduces to
\begin{equation}
    \int_{A_{\mathrm{ref}}} \delta U \,\bar{H}_0 \,\mathrm{d}A_{\mathrm{ref}}
    =
    \int_{A_{\mathrm{cur}}} \bigl(-\bar{P}\,\mathbf{n}\bigr)\cdot\delta\mathbf{u}\,\mathrm{d}A_{\mathrm{cur}}
    =-\bar{P}\,\delta V,
\end{equation}
where $\delta V$ is the variation of the volume enclosed by the deformed membrane. Rearranging,
\begin{equation}
    \int_{A_{\mathrm{ref}}} \delta U\,\bar{H}_0\,\mathrm{d}A_{\mathrm{ref}} + \bar{P}\,\delta V = 0.
\end{equation}
Adopting a spherical reference configuration of radius $R_0$ parameterised by $(\theta, \varphi)$, the reference area element is $\mathrm{d}A_{\mathrm{ref}} = R_0^{2}\sin\theta\,\mathrm{d}\theta\,\mathrm{d}\varphi$, so that
\begin{equation}
    \int_0^{2\pi}\!\!\int_0^{\pi} \delta U\,\bar{H}_0\,R_0^{2}\sin\theta\,\mathrm{d}\theta\,\mathrm{d}\varphi \;+\; \bar{P}\,\delta V \;=\; 0.
    \label{eq:vw_polar}
\end{equation}
Identifying the strain-energy density per unit reference area of the magneto-active gel membrane,
$U\equiv\bar{S}(\lambda_1, \lambda_2)$, given by \eqref{mag_gel_energy}, \eqref{eq:vw_polar} becomes
\begin{equation}
    \int_0^{2\pi}\!\!\int_0^{\pi} \delta \bar{S}\,R_0^{2}\sin\theta\,\mathrm{d}\theta\,\mathrm{d}\varphi \;+\; \bar{P}\,\delta V \;=\; 0.
\end{equation}
The following non-dimensional quantities are introduced
\begin{equation}
    S = \frac{\bar{S}}{\mu^{*}}, \,
    (\rho,\,\eta,\,a,\,H_0) = \frac{(\bar{\rho},\,\bar{\eta},\,\bar{a},\,\bar{H}_0)}{R_0}, \,
    P = \frac{\bar{P}\,R_0}{\mu^{*}\,\bar{H}_0}, \,
    C_r = \frac{\bar{C}_r\,R T}{\mu^{*}}, \,
    \nu = \frac{\bar{\nu}\,\mu^{*}}{R T}, \,
    \mu_i = \frac{\bar{\mu}_i}{\mu^{*}}.
\end{equation}
in which $\mu^{*}$ denotes the shear modulus of the dry polymer and $R T$ is the thermal energy per unit volume. With the kinematic identifications $\bar{S} = \bar{S}(\rho, \rho', \eta, \eta')$, where a prime denotes $\mathrm{d}/\mathrm{d}\theta$ operator- and calculating  the enclosed volume, $\bar{P}\,\delta V \mapsto P\bigl(\rho\,\eta'\,\delta\rho + \tfrac{1}{2}\rho^{2}\,\delta\eta'\bigr)$ per unit $\mathrm{d}\theta\,\mathrm{d}\varphi$ after non-dimensionalisation, the variation of the total potential reduces to the axisymmetric weak form
\begin{align}
    \delta\Pi=
    \int_0^{2\pi}\int_0^{\pi}\!\!\left[\,
        \left(\frac{\partial S}{\partial \rho}\,\delta\rho
        + \frac{\partial S}{\partial \rho'}\,\delta\rho'
        + \frac{\partial S}{\partial \eta}\,\delta\eta
        + \frac{\partial S}{\partial \eta'}\,\delta\eta'\right)\sin\theta
        -P\rho\,\eta'\,\delta\rho
        -\tfrac{1}{2}P\rho^{2}\,\delta\eta'
    \right]\mathrm{d}\theta\,\mathrm{d}\varphi
    =0,
    \label{eq:force_balance_wf}
\end{align}
which must hold for all kinematically admissible variations $\delta\rho$ and $\delta\eta$ satisfying the essential boundary conditions at the poles. Equation \eqref{eq:force_balance_wf} is the weak form for linear momentum balance. 

\subsection{Fluid content balance} 
Let $\bar{C}_r$ denote the number of moles of solvent absorbed by the
polymer per unit reference volume $V_0$, and let
$\mathbf{\overset{\circ}{J}}_R$ be the molar flux of solvent measured
per unit reference area and per unit time. With $\mathbf{N}_R$ the
outward unit normal to the reference surface $A_0$, the integral
$-\!\int_{A_0}\mathbf{\overset{\circ}{J}}_R\!\cdot\!\mathbf{N}_R\,
\mathrm{d}A_R$ gives the net rate at which solvent moles cross $A_0$
into $V_0$. Conservation of solvent therefore reads
\begin{align}
    \int_{V_0}\dot{\bar{C}}_r\,\mathrm{d}V_R
    \,+\, \int_{A_0}\mathbf{\overset{\circ}{J}}_R\!\cdot\!\mathbf{N}_R\,\mathrm{d}A_R
    \;=\; 0,
    \label{fluid content-1}
\end{align}
Applying the divergence theorem to the surface integral and using the
arbitrariness of the integration volume gives the local form
\begin{align}
    \dot{\bar{C}}_r \,+\, \mathrm{Div}\,\mathbf{\overset{\circ}{J}}_R
    \;=\; 0.
    \label{fluid content-2}
\end{align}

As both the polymer and the solvent are individually
incompressible, any change in the gel volume must arise from a change
in solvent content alone. This gives the swelling constraint
$J = 1 + \bar{\nu}\bar{C}_r = \phi^{-1}$, where $\phi$ is the polymer
volume fraction and $\bar{\nu}$ the molar volume of the solvent.
Differentiating in time yields
\begin{align}
    \dot{\bar{C}}_r \;=\; -\,\frac{\dot{\phi}}{\phi^{2}\,\bar{\nu}}.
    \label{Cr-dot}
\end{align}
The solvent flux is assumed to obey a Darcy-type constitutive law,
\begin{align}
    \mathbf{\overset{\circ}{J}}_R
    \;=\; -\,\frac{D}{\bar{\nu}\,R T}\,\gamma\,\mathbf{b}^{-1}\,
        \mathrm{Grad}\,\bar{\mu},
    \label{JR-consti}
\end{align}
where $D$ is the solvent diffusion coefficient, $\bar{\mu}$ is the
chemical potential of the solvent given as \citep{bouklas2015nonlinear} $\bar{\mu} = {\partial \bar{S}}/{\partial C_r}$, $\mathbf{b}$ is the right
Cauchy--Green deformation tensor, and $\gamma$ is a dimensionless
mobility factor that captures how the network ``opens up'' under
swelling. For simplicity, $\gamma$ is taken to depend linearly on the
polymer volume fraction,
\begin{equation}
    \gamma(\phi) \;=\; (1-\gamma_s)\,\phi + \gamma_s,
    \qquad \gamma_s > 1,
\end{equation}
so that $\gamma = 1$ in the dry limit ($\phi=1$) and increases above
unity as the gel swells.

Substituting \eqref{Cr-dot} and \eqref{JR-consti} into
\eqref{fluid content-2} and multiplying through by $-\bar{\nu}$ gives
\begin{align}
    \frac{\dot{\phi}}{\phi^{2}}
    \,+\, \frac{D}{R T}\,\mathrm{Div}\!\left[
        \gamma\,\mathbf{b}^{-1}\,\mathrm{Grad}\,\bar{\mu}\right]=0
    \label{phi_eq_1}
\end{align}
or, after multiplication by $\phi^{2}$,
\begin{align}
    \frac{\partial\phi}{\partial t}
    \,+\, \phi^{2}\,\frac{D}{R T}\,\mathrm{Div}\!\left[       \gamma\,\mathbf{b}^{-1}\,\mathrm{Grad}\,\bar{\mu}\right]=0
    \label{phi_eq_2}
\end{align}

We next non-dimensionalise the chemical potential and time. Introducing
$\mu = \bar{\mu}/(RT)$ and the diffusion time scale
$\tau = tD/\bar{H}_0^{2}$ (with $\bar{H}_0$ the reference mid-surface
thickness), \eqref{phi_eq_2} becomes
\begin{align}
    \frac{\partial\phi}{\partial\tau}
    \,+\, \phi^{2}\,\bar{H}_0^{2}\,\mathrm{Div}\!\left[
        \gamma\,\mathbf{b}^{-1}\,\mathrm{Grad}\,\mu\right]=0,
    \label{phi_eq_3}
\end{align}
which is the dimensionless local form of the fluid-content balance.

\subsection{Weak form}
Multiplying \eqref{phi_eq_3} by an admissible test function
$\delta\phi$ and integrating over the reference volume $V_0$ gives
\begin{align}
    \int_{V_0}\!\!\Bigl[\delta\phi\,\dot{\phi}
        \,+\,\delta\phi\,\phi^{2}\,\bar{H}_0^{2}\,
        \mathrm{Div}\!\left[\gamma\,\mathbf{b}^{-1}\,
        \mathrm{Grad}\,\mu\right]\Bigr]\,\mathrm{d}V_R=0.
    \label{weak_form_1}
\end{align}
Applying integration by parts to the divergence term moves the
divergence onto the test function and produces a surface integral on
the reference boundary $A_0$,
\begin{align}
    \int_{V_0}\!\delta\phi\,\phi^{2}\bar{H}_0^{2}\,
        \mathrm{Div}[\gamma\mathbf{b}^{-1}\mathrm{Grad}\,\mu]\,
        \mathrm{d}V_R
    \,=\,&\int_{A_0}\!\bar{H}_0^{2}\,\phi^{2}\delta\phi\,
        [\gamma\mathbf{b}^{-1}\mathrm{Grad}\,\mu]\!\cdot\!\mathbf{N}_R\,
        \mathrm{d}S_R \nonumber\\
    & -\int_{V_0}\!\bar{H}_0^{2}\,\mathrm{Grad}(\phi^{2}\delta\phi)\!
        \cdot\!\gamma\mathbf{b}^{-1}\mathrm{Grad}\,\mu\,\mathrm{d}V_R.
        \label{eq:wform}
\end{align}
Using the dimensionless form of the constitutive law
\eqref{JR-consti},
$\mathbf{\overset{\circ}{J}}_R
   = -\,(D/\bar{\nu})\,\gamma\,\mathbf{b}^{-1}\,\mathrm{Grad}\,\mu$,
together with the Robin-type surface boundary condition
\begin{equation}
    -\,\mathbf{\overset{\circ}{J}}_R\!\cdot\!\mathbf{N}_R
    \;=\; \alpha_R\bigl(\tilde{\mu} - \mu_{\infty}\bigr),
\end{equation}
where $\alpha_R = \bar{\alpha}_R\,R T$ is the non-dimensionalised
surface mass-transfer coefficient,
$\tilde{\mu} = \bar{\tilde{\mu}}/(R T)$ is the dimensionless chemical
potential at the surface, and $\mu_{\infty}$ is the chemical potential
of the surrounding solvent bath. The boundary integral term in \eqref{eq:wform} can be
written as
$$\displaystyle\int_{A_0}\!\bar{H}_0^{2}\,\phi^{2}\delta\phi\,
(\bar{\nu}/D)\,\alpha_R(\tilde{\mu}-\mu_{\infty})\,\mathrm{d}S_R.$$
Taking $\mu_{\infty}=0$ for an unbounded solvent reservoir, the weak
form of the fluid-content balance reads
\begin{align}
\label{fluid_balance_wf}
    \int_{V_0}\!\delta\phi\,\dot{\phi}\,\mathrm{d}V_R
    \,-\, \int_{V_0}\!\bar{H}_0^{2}\,
        \mathrm{Grad}(\phi^{2}\delta\phi)\!\cdot\!
        \gamma\mathbf{b}^{-1}\mathrm{Grad}\,\mu\,\mathrm{d}V_R
    \,+\, \int_{A_0}\!\bar{H}_0^{2}\,\phi^{2}\delta\phi\,
        \frac{\bar{\nu}}{D}\,\alpha_R\,\tilde{\mu}\,\mathrm{d}S_R
    \;=\;0.
\end{align}

For the thin spherical shell of reference radius $R_0$ and mid-surface
thickness $\bar{H}_0$, the volume and surface elements are
$\mathrm{d}V_R = \bar{H}_0\,R_0^{2}\sin\theta\,\mathrm{d}\theta\,\mathrm{d}\varphi$
and
$\mathrm{d}S_R = R_0^{2}\sin\theta\,\mathrm{d}\theta\,\mathrm{d}\varphi$.
Carrying out the (trivial) $\varphi$-integration by axisymmetry,
dividing through by $2\pi\,\bar{H}_0 R_0^{2}$, and expressing the
gradient operator in spherical coordinates, Eq.~\eqref{fluid_balance_wf}
reduces to
\begin{align}
\label{Diffusion_weak_form}
    \int_0^{\pi}\!\delta\phi\,\dot{\phi}\,\sin\theta\,\mathrm{d}\theta
    \;-\; & \int_0^{\pi}\!\left[H_0^{2}\,
        \frac{\partial(\phi^{2}\delta\phi)}{\partial\theta}\!\cdot\!
        \gamma\,\mathbf{b}^{-1}\,\frac{\partial\mu}{\partial\theta}\right]
        \!\sin\theta\,\mathrm{d}\theta  \nonumber\\
    \;+\; & \int_0^{\pi}\!\left[\phi^{2}\delta\phi\,H_0\,
        \frac{\bar{\nu}}{D}\,\alpha_R\,\tilde{\mu}\right]
        \!R_0\,\sin\theta\,\mathrm{d}\theta \;=\; 0.
\end{align}
The weak form of momentum balance \eqref{eq:force_balance_wf} can be rewritten as
\begin{align}
\label{Deformation_weak_form}
    \int_0^{\pi}\!\!\left[\!\left(\frac{\partial S}{\partial\rho}\delta\rho
        + \frac{\partial S}{\partial\rho'}\delta\rho'
        + \frac{\partial S}{\partial\eta}\delta\eta
        + \frac{\partial S}{\partial\eta'}\delta\eta'\right)\!\sin\theta
        \,-\, P\rho\eta'\delta\rho
        \,-\, \tfrac{P}{2}\rho^{2}\delta\eta'\right]\!\mathrm{d}\theta
    \;=\; 0.
\end{align}

Equations \eqref{Deformation_weak_form} and
\eqref{Diffusion_weak_form} together constitute the weak form of the
coupled diffusion--deformation--magnetic initial/boundary-value
problem, in which $\delta\rho$, $\delta\eta$, and $\delta\phi$ are
admissible test functions for the displacements
$({\rho},{\eta})$ and the polymer volume fraction $\phi$,
respectively.
\vspace{2mm}

\subsection{Finite element discretization}   

The coupled diffusion--deformation problem can be solved either monolithically, as in \cite{bouklas2015nonlinear}, or in a staggered (partitioned) manner, as in \cite{urrea2024comparative}. In the monolithic approach, the balance of linear momentum and the fluid-content balance are assembled into a single coupled tangent and solved simultaneously within each Newton iteration. In the staggered approach, the two balances are solved sequentially within each time step: the mechanical degrees of freedom $(\rho,\eta)$ are updated with $\phi$ held fixed, the chemical degree of freedom $\phi$ is then updated with $(\rho,\eta)$ held fixed, and the two-way coupling between the fields is enforced iteratively through an outer staggered loop described below. In this work, we adopt the staggered scheme. 
\subsubsection{Spatial discretisation}
Displacements $\rho$, $\eta$ and polymer volume fraction $\phi$ are
discretised using interpolation functions in the domain as
\begin{equation*}
    \rho = \sum U^i \rho^i,   \,\,\,\,\,\,
    \eta = \sum U^i \eta^i,   \,\,\,\,\,\,
    \phi = \sum \Phi^i \phi^i,
\end{equation*}
where $U^i$ are the displacement shape functions with nodal values
$\rho^i$ and $\eta^i$, and $\Phi^i$ are the polymer-volume-fraction shape
functions with nodal values $\phi^i$. The corresponding test functions
are
\begin{equation*}
    \delta\rho =  \delta\rho^j,   \,\,\,\,\,\,
    \delta\eta =  \delta\eta^j,   \,\,\,\,\,\,
    \delta\phi =  \delta\phi^j.
\end{equation*}

Following \citep{boger2017minimization,bouklas2015nonlinear,babuvska1971error,babuvska1973finite,brezzi2012mixed}, we adopt a Taylor--Hood-type pairing in which
$U^i$ is one polynomial order higher than $\Phi^i$, so that the LBB
condition is satisfied at the instantaneous (nearly incompressible) limit
and spurious oscillations of $\phi$ in the early-time response are
suppressed. The staggered split does not relax this requirement.

\subsubsection{Mechanical sub-problem (with \texorpdfstring{$\phi$}{phi} fixed)}

Holding $\phi$ at its most recent staggered iterate, and using the guess value $\rho$ and $\eta$ starting from dry inflation equilibrium, the residuals of the
linear momentum balance read
\begin{equation*}
    R_\rho^j = \int_0^\pi \left[ \frac{\partial S}{\partial \rho}\delta\rho^j + \frac{\partial S}{\partial \rho'} \frac{\delta\rho^j}{\partial\theta} \right]\sin{\theta}\,\text{d}\theta  - \int_0^\pi P\,\rho\,\eta'\,\delta\rho^j\,\text{d}\theta ,
\end{equation*}
\begin{equation*}
    R_\eta^j = \int_0^\pi \left[ \frac{\partial S}{\partial \eta}\delta\eta^j + \frac{\partial S}{\partial \eta'} \frac{\delta\eta^j}{\partial\theta} \right]\sin{\theta}\,\text{d}\theta  - \int_0^\pi \frac{P}{2}\,\rho^2\, \frac{\delta\eta^j}{\partial\theta}\,\text{d}\theta.
\end{equation*}
The corresponding consistent tangent blocks are
\begin{align*}
    \mathbf{K}^{jk}_{\rho\rho} = \frac{\partial R_\rho^j }{\partial \rho^k} = &\int_0^\pi \left[ \frac{\partial^2 S}{\partial \rho \partial \rho }\delta\rho^j\delta\rho^k + \frac{\partial^2 S}{\partial \rho \partial\rho' } \frac{\partial\delta\rho^j}{\partial\theta} \delta\rho^k + \frac{\partial^2 S}{\partial \rho' \partial \rho }\delta\rho^j \frac{\partial\delta\rho^k}{\partial \theta} + \frac{\partial^2 S}{\partial \rho' \partial\rho' } \frac{\partial\delta\rho^j}{\partial\theta} \frac{ \partial \delta\rho^k}{\partial \theta}     \right]\sin{\theta}\,\text{d}\theta \\
    &- \int_0^\pi P\,\eta'\,\delta\rho^j\,\delta\rho^k \,\text{d}\theta ,
\end{align*}
\begin{align*}
    \mathbf{K}^{jk}_{\rho\eta} =\frac{\partial R_\rho^j}{\partial \eta^k} =  &\int_0^\pi \left[ \frac{\partial^2 S}{\partial \eta \partial \rho }\delta\rho^j\delta\eta^k + \frac{\partial^2 S}{\partial \eta \partial\rho' } \frac{\partial\delta\rho^j}{\partial\theta} \delta\eta^k + \frac{\partial^2 S}{\partial \eta' \partial \rho }\delta\rho^j \frac{\partial\delta\eta^k}{\partial \theta} + \frac{\partial^2 S}{\partial \eta' \partial\rho' } \frac{\partial\delta\rho^j}{\partial\theta} \frac{ \partial \delta\eta^k}{\partial \theta}     \right]\sin{\theta}\,\text{d}\theta \\
    &- \int_0^\pi P\,\delta\rho^j\,\frac{\partial\delta\eta^k}{\partial\theta} \,\text{d}\theta ,
\end{align*}
\begin{align*}
    \mathbf{K}^{jk}_{\eta\rho} =\frac{\partial R_\eta^j}{\partial \rho^k} =&\int_0^\pi \left[ \frac{\partial^2 S}{\partial \rho \partial \eta }\delta\eta^j\delta\rho^k + \frac{\partial^2 S}{\partial \rho \partial\eta' } \frac{\partial\delta\eta^j}{\partial\theta} \delta\rho^k + \frac{\partial^2 S}{\partial \rho' \partial \eta }\delta\eta^j \frac{\partial\delta\rho^k}{\partial \theta} + \frac{\partial^2 S}{\partial \rho' \partial\eta' } \frac{\partial\delta\eta^j}{\partial\theta} \frac{ \partial \delta\rho^k}{\partial \theta}     \right]\sin{\theta}\,\text{d}\theta \\
    &- \int_0^\pi P\,\rho\,\frac{\partial \eta^j}{\partial \theta} \delta\rho^k  \text{d}\theta ,
\end{align*}
\begin{align*}
    \mathbf{K}^{jk}_{\eta\eta} = \frac{\partial R_\eta^j}{\partial \eta^k}=&\int_0^\pi \left[ \frac{\partial^2 S}{\partial \eta \partial \eta }\delta\eta^j\delta\eta^k + \frac{\partial^2 S}{\partial \eta \partial\eta' } \frac{\partial\delta\eta^j}{\partial\theta} \delta\eta^k + \frac{\partial^2 S}{\partial \eta' \partial \eta }\delta\eta^j \frac{\partial\delta\eta^k}{\partial \theta} + \frac{\partial^2 S}{\partial \eta' \partial\eta' } \frac{\partial\delta\eta^j}{\partial\theta} \frac{ \partial \delta\eta^k}{\partial \theta}     \right]\sin{\theta}\,\text{d}\theta.
\end{align*}

\subsubsection{Diffusion sub-problem (with \texorpdfstring{$\rho$}{ρ} and \texorpdfstring{$\eta$}{η} fixed)}

Time integration of the fluid-content balance is carried out with a
backward-Euler scheme, with $\phi^n$ denoting the converged volume
fraction at the previous time step. Holding the deformation
$(\rho,\eta)$ fixed at the value just produced by the mechanical
sub-problem, the residual reads
\begin{align*}
    R_\phi^j =  \int_0^\pi  \,\delta\phi^j\,\frac{ \phi - \phi^n }{\Delta \tau} \,\sin{\theta}\,\text{d}\theta &-\int_0^{\pi} \left[ H_o^2 \frac{\partial}{\partial \theta}(\phi^2\,\delta\phi^j).\gamma\mathbf{b^{-1}}\frac{\partial}{\partial\theta}(\mu)  \right] \sin{\theta}\,\text{d}\theta\   \\ \nonumber
    &+\int_0^\pi\left[  \phi^2 \,\delta\phi^j\,H_o \frac{\bar{\nu}}{D}  \alpha_R( \tilde{\mu} )   \right] R_o\,\sin{\theta}\,\text{d}\theta\
\end{align*}
or, after multiplying through by $\Delta\tau$ and expanding the spatial
derivative,
\begin{align*}
    R_\phi^j =  \int_0^\pi  \,\delta\phi^j\,({ \phi - \phi^n })\,\sin{\theta}\,\text{d}\theta &-\Delta \tau\int_0^{\pi} \left[ H_o^2 \,\left(2\,\phi\,\phi'\,\delta\phi^j + \phi^2 \frac{\partial \delta\phi^j }{\partial\theta} \right).\gamma\mathbf{b^{-1}}\frac{\partial}{\partial\theta}(\mu)  \right] \sin{\theta}\,\text{d}\theta\   \\ \nonumber
    &+\Delta \tau\int_0^\pi\left[  \phi^2 \,\delta\phi^j\,H_o \frac{\bar{\nu}}{D}  \alpha_R( \tilde{\mu} )   \right] R_o\,\sin{\theta}\,\text{d}\theta\
\end{align*}
The corresponding tangent block is
\begin{align*}
    &\mathbf{K}_{\phi\phi}^{jk} =  \frac{\partial R_\phi^j}{\partial \phi^k} =\int_0^\pi \delta\phi^j\,\delta\phi^k\,\sin{\theta}\,\text{d}\theta - \Delta\tau\,H_o^2 \int_0^\pi  \biggl\{ \left[ 2\phi'\delta\phi^j\delta\phi^k + 2\phi\,\delta\phi^j \frac{\partial \delta \phi^k}{\partial \theta} \right.   \\ \nonumber
    & \left.\qquad\qquad\qquad+ 2\phi \frac{\partial \delta\phi^j}{\partial\theta}\delta\phi^k  \right].\gamma\mathbf{b}^{-1}\frac{\partial \mu}{\partial \theta}+ \left[ 2\phi\phi'\delta\phi^j + \phi^2 \frac{\partial \delta \phi ^j}{\partial \theta} \right].\gamma\mathbf{b}^{-1} \frac{\partial^2 \mu }{\partial \phi \partial \theta} \delta\phi^k \biggl\}\sin{\theta}\,\text{d}\theta +   \\ \nonumber
    &\Delta\tau\int_0^\pi \left[ 2\phi\,\delta\phi^k\,\delta\phi^j\,H_o\,\frac{\bar{\nu}}{D} \alpha_R(\bar{\mu})\right]R_o\,\sin{\theta}\,\text{d}\theta+\Delta\tau \int_0^\pi \left[ \phi^2 \,\delta\phi^j\,H_o \frac{\bar{\nu}}{D}  \alpha_R \frac{ \partial \tilde{\mu}}{\partial \phi}\delta\phi^k \right]R_o\sin{\theta}\,\text{d}\theta.
\end{align*}

\subsubsection{Staggered solution algorithm}

Within each time step $[\tau^n,\,\tau^{n+1}=\tau^n+\Delta\tau]$ the
solution is advanced through an outer staggered loop indexed by~$k$.
At the start of the time step we initialise
$(\rho^{(0)},\eta^{(0)},\phi^{(0)})\!=\!(\rho^n,\eta^n,\phi^n)$.
Each outer iteration consists of the following two steps.

\textbf{Step~1 -- Mechanical sub-problem.}
With $\phi=\phi^{(k)}$ frozen, update $(\rho,\eta)$ by Newton--Raphson:
\begin{align}
    \begin{bmatrix}
         \mathbf{K}_{\rho\rho} & \mathbf{K}_{\rho\eta} \\
         \mathbf{K}_{\eta\rho} & \mathbf{K}_{\eta\eta}
    \end{bmatrix} \begin{bmatrix}
         \Delta\rho \\ \Delta \eta
    \end{bmatrix}   =   \begin{bmatrix}
          R_\rho \\ R_\eta
    \end{bmatrix},
\end{align}
iterated until $\|[R_\rho,\,R_\eta]\|$ falls below the Newton tolerance,
giving the updated displacements
$(\rho^{(k+1)},\eta^{(k+1)})$ for the frozen value of $\phi^{(k)}$.

\textbf{Step~2 -- Diffusion sub-problem.}
With $(\rho,\eta)=(\rho^{(k+1)},\eta^{(k+1)})$ frozen, update $\phi$ by
Newton--Raphson:
\begin{align}
    \mathbf{K}_{\phi\phi}\,\Delta\phi = R_\phi,
\end{align}
iterated until $\|R_\phi\|$ falls below the Newton tolerance, giving the
updated polymer volume fraction $\phi^{(k+1)}$.

\textbf{Outer convergence check.}
The outer staggered loop repeats Steps~1--2 until the changes in both
fields between successive outer iterates fall below prescribed tolerances,
\begin{equation*}
    \frac{\bigl\|(\rho^{(k+1)},\eta^{(k+1)}) - (\rho^{(k)},\eta^{(k)})\bigr\|}
         {\bigl\|(\rho^{(k+1)},\eta^{(k+1)})\bigr\|} < \mathrm{tol}_u,
    \qquad
    \frac{\bigl\|\phi^{(k+1)} - \phi^{(k)}\bigr\|}{\bigl\|\phi^{(k+1)}\bigr\|}
    < \mathrm{tol}_\phi,
\end{equation*}
at which point the converged values are accepted as the solution at
$\tau^{n+1}$, i.e.\
$(\rho^{n+1},\eta^{n+1},\phi^{n+1}) \leftarrow (\rho^{(k+1)},\eta^{(k+1)},\phi^{(k+1)})$,
and the algorithm advances to the next time step.

Because the mechanical and diffusion sub-problems are  solved separately, each time step requires three nested convergence checks, one for the mechanical Newton loop, one for the diffusion Newton loop, and one for the outer staggered loop, in contrast to the
single convergence check of the monolithic Newton scheme. The trade-off is two smaller, better-conditioned linear systems per outer iteration and the ability to use independently tuned solvers for the mechanical and diffusion blocks. The staggered, fully coupled nonlinear transient formulation described above is
implemented in the FEniCS Project \cite{logg2012automated, alnaes2015fenics},
an open-source Python framework for the automated solution of partial
differential equations by the finite element method.

\section{Numerical results}
\label{sec: result}
In this section, the numerical simulation results are presented in three stages. We first consider the
free swelling of a gel balloon, in the absence of applied pressure and a magnetic field. We
then examine the inflation of the balloon in the absence of a magnetic field,
and finally the inflated balloon under an applied magnetic field. For the
inflated balloon, we study two limiting equilibrium responses. In
\emph{dry inflation equilibrium}, the balloon is never exposed to the solvent, so there is
no solvent uptake. In the \emph{final equilibrium configuration}, the balloon is immersed in the
solvent and the pressure is increased slowly compared with the time scale of
solvent diffusion, so that the balloon is fully swollen at every point on the
equilibrium curve. 
\subsection{Response of a gel in the absence of magnetic field}
\label{sec: absence of magnetic field}   

\subsubsection{Free swelling} \label{sec: free_swelling}
When a dry, undeformed polymeric balloon is immersed in water, it creates a difference in chemical potential, i.e., a higher chemical potential of the surrounding solvent and a lower chemical potential inside the dry polymer. This difference in chemical potential produces an osmotic pressure. This pressure pushes solvent molecules inside the dry polymer membrane of the balloon, making it swell.

Two forces are in play: the osmotic pressure, which stretches and swells the polymeric membrane due to the entry of solvent molecules, and the cross-linked polymeric chains, which resist the swelling. Therefore, equilibrium is attained at a stress-free swollen state, where the resistance from the cross-linked polymeric chains balances the osmotic pressure. Hence, the free-swollen equilibrium state is associated with a stress-free configuration and isotropic stretch $\lambda_\theta = \lambda_\phi = \lambda_r = \lambda_0 $. Since the free swelling state is a stress free configuration, $\lambda_0$ can be determined by equating the Cauchy stress of the hydrogel to zero. For a given value of $\nu$, $\lambda_0$ can be determined as,
\begin{align}
    &\kappa \left[\frac{3}{\beta \lambda_0{\left(\theta \right)}} - \frac{3 \left(\lambda_0^{3}{\left(\theta \right)}\right)^{- \beta}}{\beta \lambda_0{\left(\theta \right)}}\right] + \frac{1}{{\lambda_0^{4}{\left(\theta \right)}}}  \left[{3 \mu_{1} \lambda_0^{\alpha_{1} + 3}{\left(\theta \right)} + 3 \mu_{2} \lambda_0^{\alpha_{2} + 3}{\left(\theta \right)} + 3 \mu_{3} \lambda_0^{\alpha_{3} + 3}{\left(\theta \right)}}\right] \nonumber  \\  & + \frac{1}{\nu}\left[\frac{3 \chi}{\lambda_0^{4}{\left(\theta \right)}} + 3 \lambda_0^{2}{\left(\theta \right)} \log{\left(1 - \frac{1}{\lambda_0^{3}{\left(\theta \right)}} \right)} + \frac{3}{\lambda_0{\left(\theta \right)}}\right] = 0.
    \label{ISO_SWELL}
\end{align}
For a value of $\nu=0.02744$ in equation \eqref{ISO_SWELL} gives $\lambda_0=1.5$, which leads to volume fraction $\phi=\frac{1}{J}=\frac{1}{\lambda_0^3} = 0.2963$.

\begin{figure}[H]
    \centering
    \includegraphics[width=\linewidth]{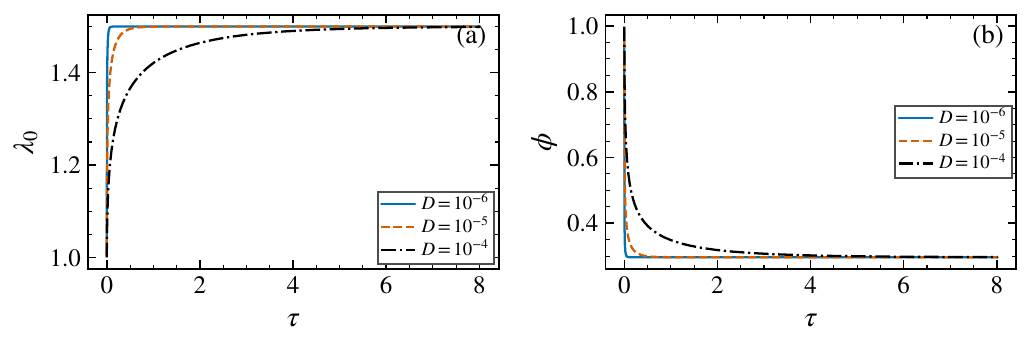}
   \caption{Free-swelling time evolution of (a) principal stretch~$\lambda_0$ and (b) volume fraction~$\phi$ for $D = 10^{-6}~\text{m}^2/\text{sec} ,\,10^{-5}~\text{m}^2/\text{sec},\,10^{-4}\text{m}^2/~\text{sec}$.}
    \label{fig:ISO_swell}
\end{figure}

The weak forms \eqref{Diffusion_weak_form} and \eqref{Deformation_weak_form}  are solved using a staggered finite element approach with $\nu = 0.02744$. The solvent diffusivity $D$ is assumed to be constant, isotropic, and independent of the deformation gradient $\mathbf{F}$ and concentration $C_r$. The time--dependent FEM results for $\tau$ vs.\ $\lambda_0$ and $\tau$ vs.\ $\phi$ are shown in Fig.~\ref{fig:ISO_swell}. The free swelling process of dry polymeric balloon is simulated giving the equilibrium isotropic stretch $\lambda_0 = 1.5$ is shown in Fig.~\ref{fig:ISO_swell}(a), and $\phi = 0.2963$ in Fig.~\ref{fig:ISO_swell}(b), as obtained by solving equation \eqref{ISO_SWELL}. The diffusivity $D$ governs the time required for the diffusion process to reach equilibrium.

\subsubsection{Delayed Burst} \label{sec: Delayed_burst}
The delayed burst phenomenon in thin and thick spherical gel balloons has been reported in \cite{TengLi_JMPS2019}, and in toroidal gel balloons in \cite{TAMADAPU2022}. Following these works, we assume that the time-dependent response of the hydrogel arises solely from solvent transport through the cross-linked polymer network, neglecting viscoelastic effects. Under suddenly applied mechanical loads, the hydrogel exhibits an initial incompressible deformation, followed by a long-term time-dependent evolution toward equilibrium driven by solvent diffusion.

\begin{figure}[H]
    \centering
    \includegraphics[width=0.5\linewidth]{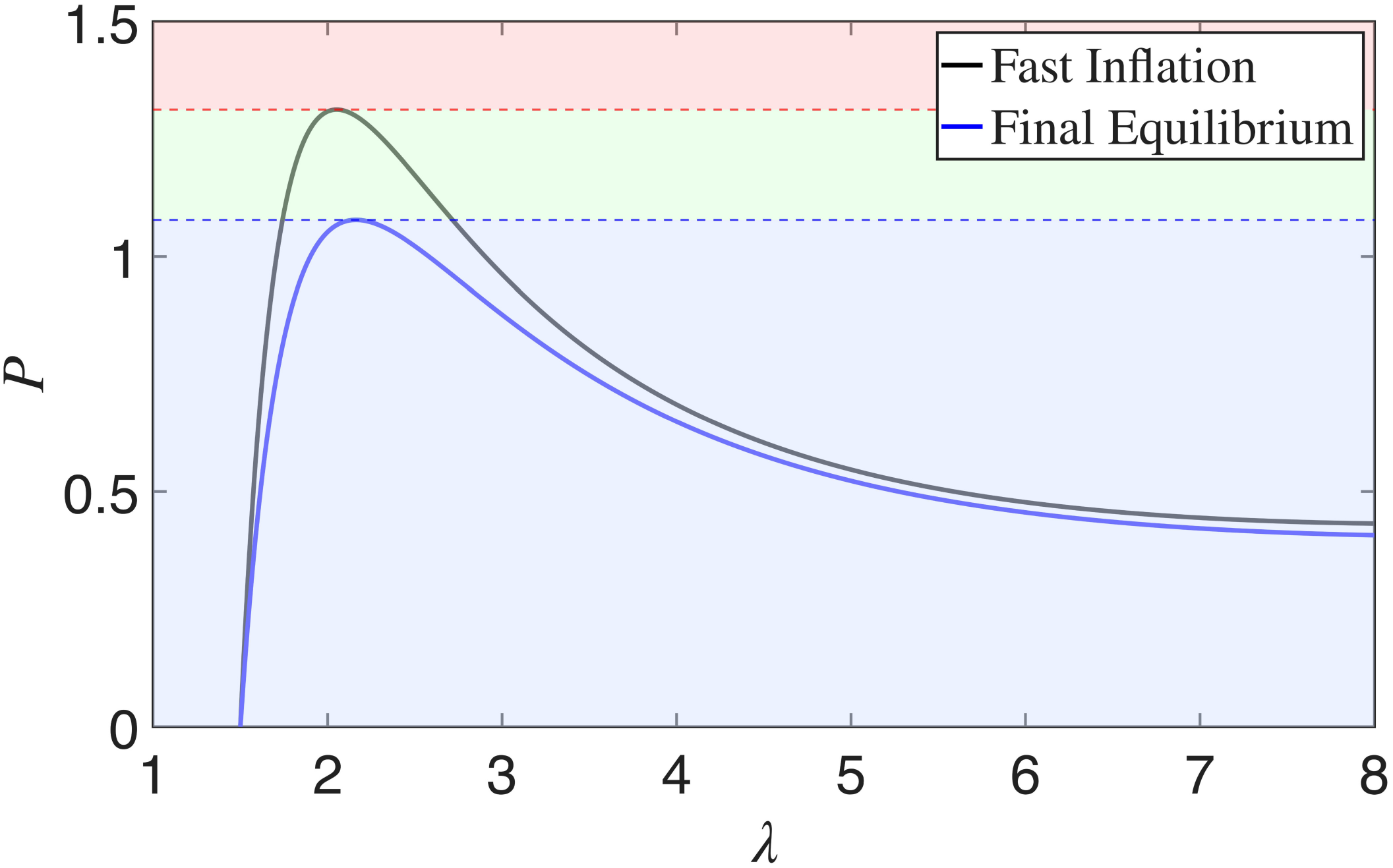}
    \caption{$P$ vs $\lambda$ for fast inflation and final swollen equilibrium configuration of gel balloon.}
    \label{fig:L2_vs_P_delayed}
\end{figure}

When the free-swollen balloon described in section (\ref{sec: free_swelling}) is subjected to rapid pressurization, it first undergoes fast inflation and then evolves toward the slow-inflation equilibrium when the pressure is held constant, as shown in Fig.~\ref{fig:L2_vs_P_delayed}. For pressures below the critical slow-inflation pressure (blue region), there is a stable equilibrium. For pressures between the critical slow and fast-inflation pressures (green region), the gel balloon undergoes a delayed burst and an instant burst in the red region, as determined using \textbf{Shooting Method} in the Appendix (\ref{sec:Appendix})  and reported in \cite{TengLi_JMPS2019}.

\begin{figure}[H]
    \centering
    \includegraphics[width=\linewidth]{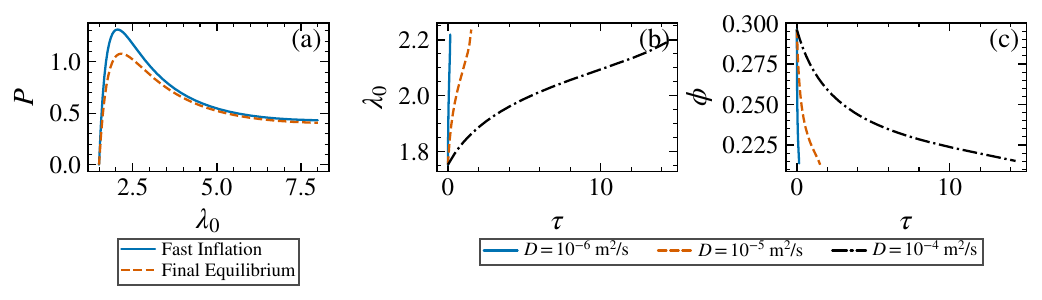}
    \caption{Delayed-burst dynamics at $P = 1.1$. (a)~$P$ vs.\ $\lambda_0$ for the fast- and slow-inflation branches. Time evolution of (b)~$\lambda_0$ and (c)~$\phi$ for $D = 10^{-6} \text{m/s},\,10^{-5}\text{m/s},\,10^{-4}\text{m/s}$.}
    \label{fig:Delayed_burst}
\end{figure}

Using the staggered FEM formulation (section (\ref{Method_II})) in this delayed-burst regime, the time evolution of the stretch $\lambda_\theta$ and the polymer fraction $\phi$ are shown in Fig.~\ref{fig:Delayed_burst}(b) and \ref{fig:Delayed_burst}(c), respectively. The results indicate continuous, uncontrolled inflation with increasing $\lambda_\theta$ (Fig.~\ref{fig:Delayed_burst}(b)) and progressive solvent uptake reflected by decreasing $\phi$ (Fig.~\ref{fig:Delayed_burst}(c)). The simulation ultimately fails to reach a converged equilibrium configuration, demonstrating the delayed burst region obtained by and  consistent with the $\textbf{Shooting Method}$ reported in the Appendix \ref{sec:Appendix}.

The consistency of the FEM results for free swelling (section \ref{sec: free_swelling}) and delayed burst (section \ref{sec: Delayed_burst}) with analytically tractable solutions and previously reported studies demonstrates the correctness and reliability of the proposed finite element formulation and numerical implementation.

\subsubsection{Swelling of pressurized balloons} \label{sec: Swelling of pressurized balloons}
Consider an inflated polymeric balloon in its completely dehydrated state ($C_r = 0$), subjected to a pressure lower than the critical pressure of the slow-inflation curve. In this section, we present FEM simulations for the evolution of an inflated dry polymer balloon immersed in solvent, showing diffusion-driven swelling until equilibrium is reached.

\begin{figure}[H]
    \centering
    \includegraphics[width=\linewidth]{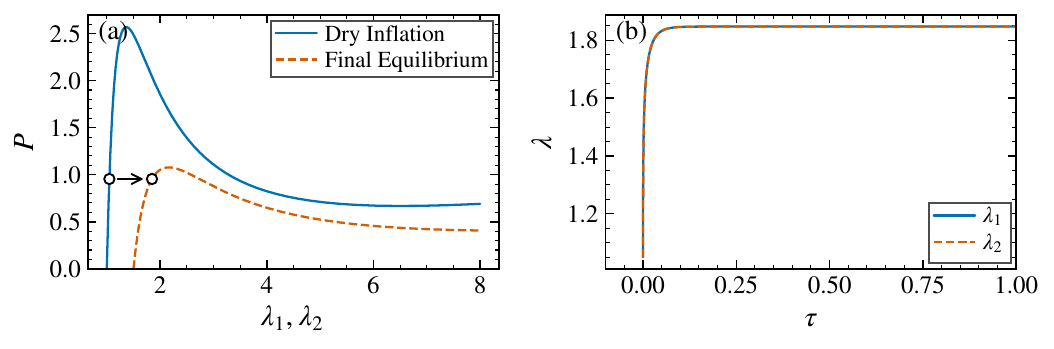}
    \caption{(a)~$P$ vs.\ $\lambda_1$, $\lambda_2$ for the dry inflation and final equilibrium branches. (b) Time evolution meridional and circumferential stretch, $\lambda_1$ and $\lambda_2$.} 
    \label{fig:Equator_pole1_sym}
\end{figure}

The equilibrium configuration of the dry polymeric balloon at pressure \( P = 0.95445 \) and stretch \( \lambda_2 = 1.04830 \) is shown in Fig.~\ref{fig:Equator_pole1_sym}(a). Upon immersion in solvent, the balloon swells and evolves toward the slow-inflation equilibrium at the same pressure, corresponding to a swollen stretch \( \lambda_2 = 1.8471 \), as indicated by the slow-inflation curve. The time evolution of the stretches at the equator and poles, shown in Fig.~\ref{fig:Equator_pole1_sym}(b)--\ref{fig:Equator_pole1_sym}(d), demonstrates that both \( \lambda_2 \) and \( \lambda_1 \) increase monotonically and converge to \( \lambda_2 = \lambda_1 = 1.8471 \). This behavior indicates a uniform swelling of the spherical balloon, preserving its spherical shape, and ultimately reaching the spherical equilibrium configuration on the slow-inflation curve, consistent with the results obtained by using \textbf{Shooting Method} (\ref{sec:Appendix})) and  reported in \cite{TengLi_JMPS2019}.

\begin{figure}[H]
    \centering
    \includegraphics[width=\linewidth]{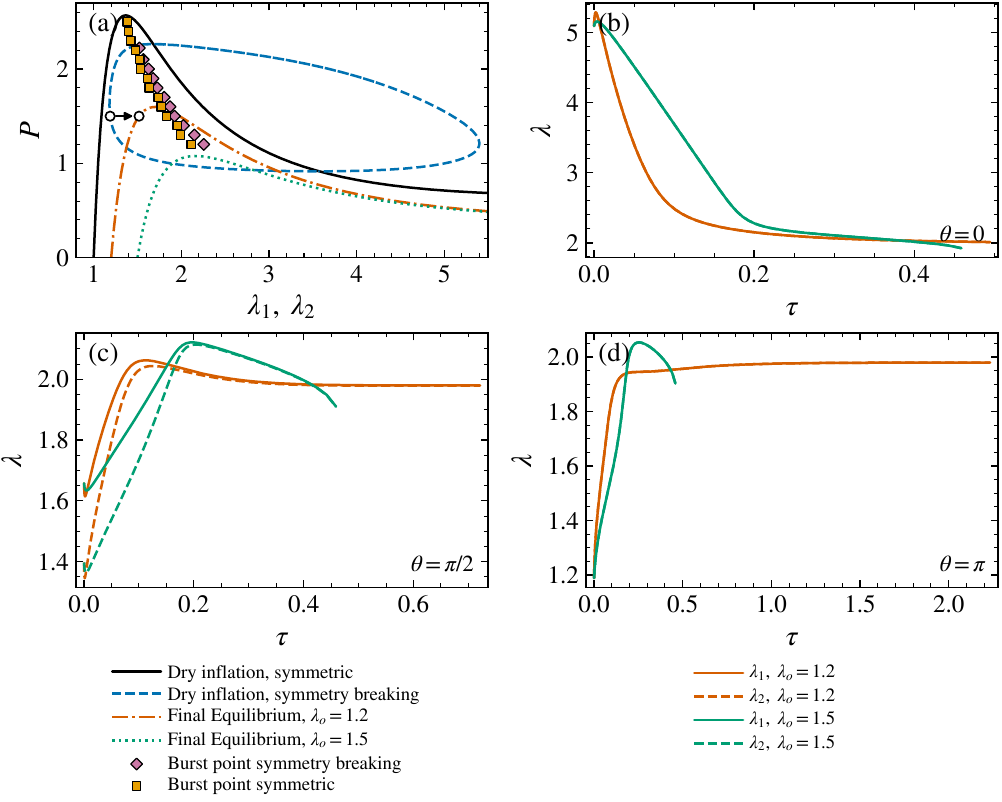}
    \caption{Swelling from a dry symmetry-breaking configuration onto the symmetric final equilibrium branch. (a)~$P$ vs.\ $\lambda_1$, $\lambda_2$ at the pole. Time evolution of $\lambda_1$ and $\lambda_2$ (b)~at the equator, (c)~at the pole $\theta = 0$, and (d)~at $\theta = \pi$.}
    \label{fig:Equator_pole1}
\end{figure}

We next consider a dry balloon held at a symmetry-breaking (pear-shaped)
equilibrium at $P = 1.5$, which is then immersed in a solvent bath while the
pressure is held constant. Figure~\ref{fig:Equator_pole1}(a) shows the dry
equilibrium curves for the symmetric and symmetry-breaking branches. It also
shows the equilibrium curves reached at the end of the diffusion process
(final equilibrium curves)  for $\lambda_0 = 1.2$
(dash-dotted) and $\lambda_0 = 1.5$ (dotted). Each value of $\lambda_0$, the
isotropic swelling stretch, corresponds to a different solvent molar volume
$\nu$. Figures~\ref{fig:Equator_pole1}(b)--\ref{fig:Equator_pole1}(d) show how
the principal stretches evolve in time at the poles ($\theta = 0$ and
$\theta = \pi$) and at the equator ($\theta = \pi/2$) for both values of
$\lambda_0$. In both cases, the balloon starts from the same dry pear-shaped
configuration, which is strongly asymmetric: the stretch at the pole
$\theta = 0$ ($\lambda_1=\lambda_2 = 5.2$) is roughly four times that at
$\theta = \pi$ ($\lambda_1=\lambda_2 = 1.25$).

Immediately after the balloon is exposed to outside solvent, the stretches undergo a rapid, short-lived
adjustment, visible as the sharp changes near $\tau = 0$ in
Figs.~\ref{fig:Equator_pole1}(b)--\ref{fig:Equator_pole1}(d), through which the balloon passes through
different pear-shaped configurations. Throughout this stage the two principal
stretches remain equal at the poles, as required by axisymmetry, but differ at
the equator. Subsequently, the stretch at $\theta = 0$ decreases monotonically,
whereas the stretch at $\theta = \pi$ increases, so the stretches at the two
poles approach each other. For $\lambda_0 = 1.2$, the prescribed pressure lies
below the limit-point pressure of the corresponding final equilibrium curve
($P_{\max} = 1.6$), and an equilibrium state therefore exists. The
equatorial stretches increase rapidly, reaching a maximum of about $2.06$ at
$\tau = 0.1$, and then decrease. All stretches converge to a common value,
$\lambda_1 = \lambda_2 = 1.98$, and the balloon attains a spherical
equilibrium configuration on the slow-inflation curve. For $\lambda_0 = 1.5$, by
contrast, $P = 1.5$ exceeds the maximum pressure the swollen balloon can sustain
($P_{\max} = 1.08$), so no equilibrium state is available. The stretches
at the $\theta = \pi/2$ and at $\theta = \pi$ reach a maximum at
$\tau = 0.2$--$0.25$ and then decrease as shown in Figs.~\ref{fig:Equator_pole1}(c) and \ref{fig:Equator_pole1}(d), and the computation terminates at
burst at $\tau = 0.46$. At that instant the stretches at all three
locations ($\theta = 0 , \, \pi/2 , \, \pi$.) are nearly equal ($\lambda_1 = \lambda_2 = 1.9$).

Figure~\ref{fig:Equator_pole1}(a) also shows the burst points obtained with
$\lambda_0 = 1.5$ ($\nu = 0.00068$) for dry balloons initially in either a
symmetric (squares) or a symmetry-breaking (diamonds) configuration over a range
of pressures. Each burst point marks the stretch at the pole $\theta = 0$, where
burst occurs, at the instant of burst. Burst points for the symmetry-breaking
case are reported only for $P \leq 2.25$, since the dry symmetry-breaking
branch does not exist above this pressure. For both symmetric and symmetry breaking initial configurations, the
stretch at burst increases as the pressure decreases. The time to burst also
increases as the pressure decreases (not shown). At a given pressure, a balloon
starting from a symmetry-breaking configuration reaches a larger polar stretch
at burst than one starting from a symmetric configuration, yet it bursts
earlier. In other words, a symmetric dry configuration survives longer in the
bath than a pear-shaped one at the same pressure.

\begin{figure}[H]
    \centering
    \includegraphics[width=\linewidth]{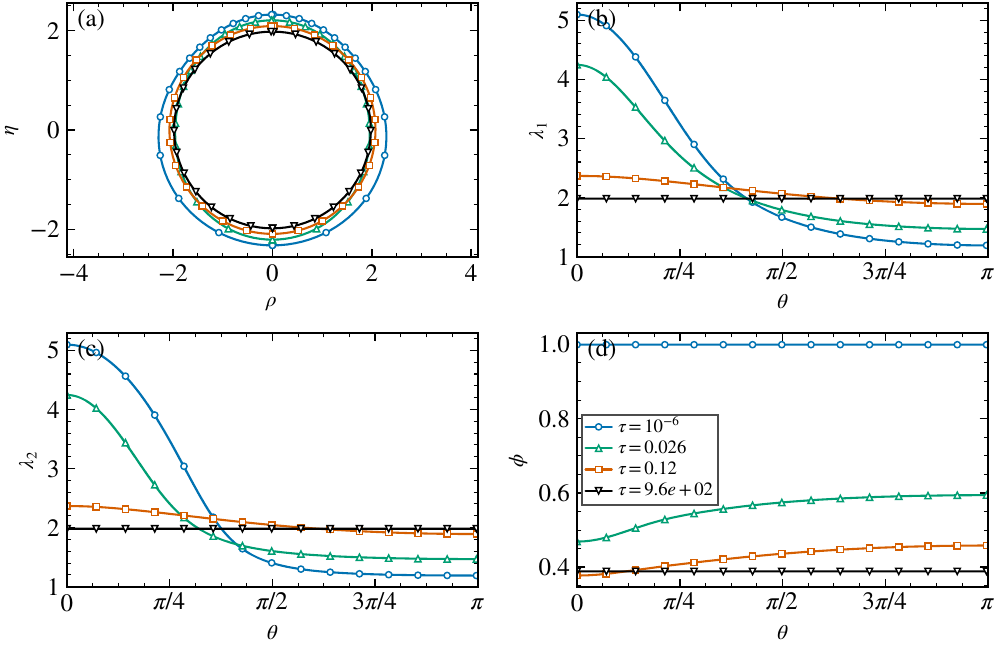}
    \caption{Time evolution of a dry symmetry-breaking configuration swelling at
    constant pressure $P = 1.5$ with $\lambda_0 = 1.2$:
    (a)~deformed shape; (b)~meridional stretch $\lambda_1(\theta)$; (c)~circumferential stretch $\lambda_2(\theta)$;
    (d)~polymer volume fraction $\phi(\theta)$.}
    \label{fig:TIME_steps}
\end{figure}

Figure~\ref{fig:TIME_steps} shows the time evolution of a dry symmetry-breaking
configuration immersed in a solvent bath with $\lambda_0 = 1.2$. The deformed
shapes at successive times are shown in Fig.~\ref{fig:TIME_steps}(a), and the
corresponding distributions of the stretches $\lambda_1$ and $\lambda_2$ along
the meridian are shown in Figs.~\ref{fig:TIME_steps}(b) and \ref{fig:TIME_steps}(c), respectively.
Immediately after the balloon is exposed to outside solvent ($\tau = 10^{-6}$), the stretches are strongly
non-uniform: $\lambda_1 = \lambda_2 = 5.1$ at the pole $\theta = 0$,
whereas $\lambda_1 = \lambda_2 = 1.2$ at the pole $\theta = \pi$,
confirming that the initial dry configuration is symmetry-breaking. Within a
very short time ($\tau = 0.026$), the balloon transitions to a different
symmetry-breaking configuration, in which the stretch at $\theta = 0$ has
dropped to about $4.25$ and that at $\theta = \pi$ has risen to about $1.5$. As
swelling proceeds, the stretch distributions flatten ($\tau = 0.12$), and the
balloon eventually attains the spherical configuration with uniform stretches
$\lambda_1 = \lambda_2 = 1.98$. The corresponding evolution of the
polymer volume fraction $\phi$ is shown in Fig.~\ref{fig:TIME_steps}(d).
Initially $\phi = 1$ everywhere, which corresponds to the dry state. As solvent is
absorbed, $\phi$ decreases non-uniformly, being lower near $\theta = 0$, where
the stretches are larger. It finally reaches a uniform equilibrium value of
$\phi = 0.39$.

\begin{figure}[H]
    \centering
    \includegraphics[width=\linewidth]{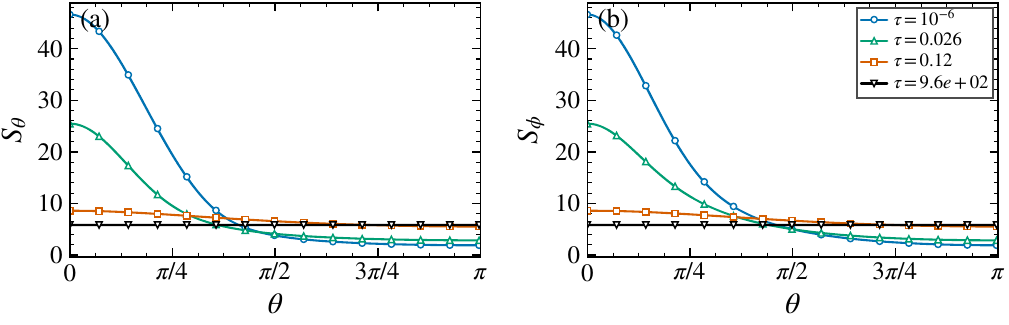}
    \caption{Time evolution of the principal stresses for the swelling process
    shown in Fig.~\ref{fig:TIME_steps} ($P = 1.5$, $\lambda_0 = 1.2$):
    (a)~meridional stress $S_\theta(\theta)$; (b)~circumferential stress $S_\phi(\theta)$.}
    \label{fig:ISO_swell_1}
\end{figure}

The corresponding evolution of the principal stresses $S_\theta$ (meridional) and $S_\phi$ (circumferential) is
shown in Fig.~\ref{fig:ISO_swell_1}. Immediately after the balloon is exposed to outside solvent
($\tau = 10^{-6}$), the stress distribution is highly non-uniform and closely
follows the stretch distribution in Figs.~\ref{fig:TIME_steps}(b) and \ref{fig:TIME_steps}(c). The
stresses reach about $47$ at the pole $\theta = 0$ and fall to about $2$ at
$\theta = \pi$. The two principal stresses are equal at both poles, as required
by axisymmetry, and differ elsewhere along the meridian. As swelling proceeds,
the stress at $\theta = 0$ decreases rapidly, to about $25$ at $\tau = 0.026$
and $8.6$ at $\tau = 0.12$, while the stress near $\theta = \pi$ increases. At
equilibrium, both stresses become uniform and equal,
$S_\theta = S_\phi = 5.9$, consistent with the spherical configuration.
Swelling therefore removes the strong stress concentration at the highly
stretched pole of the initial symmetry-breaking configuration, and the stress
becomes uniform over the membrane.

\subsection{Response of gel in the presence of external magnetic field}
\label{sec: Magnetic} 
\subsubsection{Symmetric magnetic loading}

\label{sec:Symmetric magnetic loading}
In this section, we study the swelling of an inflated balloon subjected to the
magnetic field of a coil of radius $a = 2$ placed in the equatorial plane of the
balloon. The solvent parameter is $\nu = 0.10668$, which corresponds to an
isotropic free-swelling stretch of $\lambda_0 = 1.2$. Throughout, we compare two
solution approaches: \textbf{Shooting Method}, the equilibrium energy-minimization
procedure described in section~\ref{sec:Appendix}, and \textbf{FEM}, the
coupled diffusion--deformation finite-element formulation described in
section~\ref{Method_II}.

\begin{figure}[H]  
    \centering
    \includegraphics[width=\linewidth]{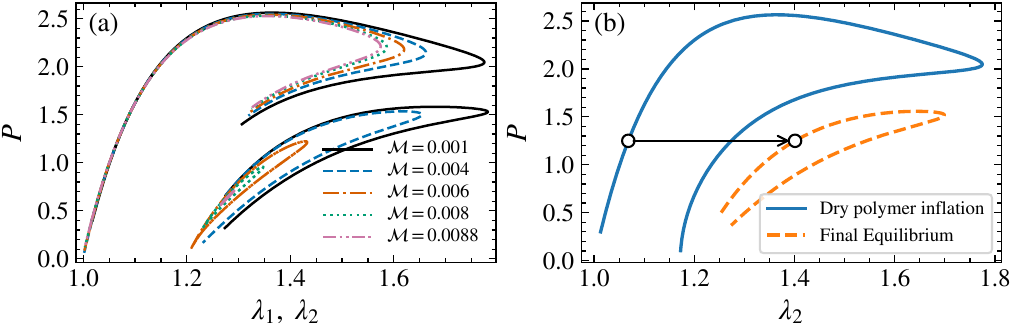}
    
    \caption{(a) Pressure--stretch curves at the pole ($\lambda_1 = \lambda_2$) for
    dry inflation equilibrium configurations and final equilibrium configurations, for various values of $\mathcal{M}$. (b) Pressure--stretch curves ($P$ vs.\ $\lambda_2$ at the pole) at $\mathcal{M} = 0.001$, $a = 2$: dry inflation (solid blue) and equilibrium swelling via \textbf{Shooting Method} (dashed orange). The arrow indicates the swelling process at $P = 1.2487$.}
    \label{fig:P_vs_lambda_Dry_Slow_pole}
\end{figure}

\begin{figure}[H]
    \centering
    \includegraphics[width=1\linewidth]{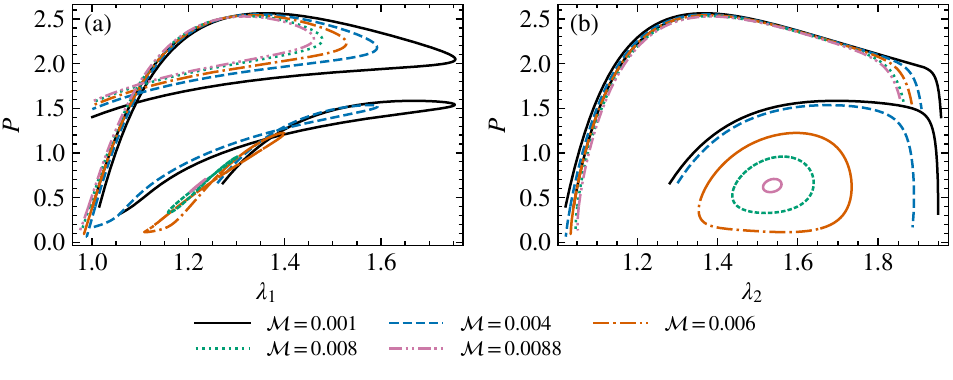}
    \caption{Pressure--stretch curves at the equator for dry inflation equilibrium and final equilibrium configuration,
    for various values of $\mathcal{M}$: (a)~$P$ vs.\ $\lambda_1$;
    (b)~$P$ vs.\ $\lambda_2$.}
    \label{fig:P_vs_lambdas_Dry_Slow_equator}
\end{figure}

Figure~\ref{fig:P_vs_lambda_Dry_Slow_pole}(a) shows the pressure--stretch curves at
the pole, where $\lambda_1 = \lambda_2$, for dry equilibrium as well as for final equilibrium configurations and for several values of the magnetic field intensity $\mathcal{M}$. The corresponding curves at the equator, $P$ vs.\ $\lambda_1$ and $P$ vs.\ $\lambda_2$, are shown in
Figs.~\ref{fig:P_vs_lambdas_Dry_Slow_equator}(a) and \ref{fig:P_vs_lambdas_Dry_Slow_equator}(b), respectively.       

The magnetic field has a pronounced effect on the slow-inflation response. For
$\mathcal{M} \geq 0.006$, the slow-inflation equilibrium curve closes on itself
and forms an isola-like closed loop that does not reach $P = 0$. This loop
shrinks as $\mathcal{M}$ increases. At the pole, the loop spans the pressure
range $0.11 \leq P \leq 1.22$ for $\mathcal{M} = 0.006$, but only
$0.56 \leq P \leq 0.71$ for $\mathcal{M} = 0.0088$. Consequently, for
these field intensities, swollen equilibrium configurations exist only within a
bounded window of pressure.

Next, we hold the pressure fixed at $P = 1.2487$ and follow the swelling of the inflated balloon for two field intensities, $\mathcal{M} = 0.001$. Fig.~\ref{fig:P_vs_lambda_Dry_Slow_pole}(b) shows the pole pressure--stretch curves at $\mathcal{M} = 0.001$ computed with \textbf{Shooting Method}, for dry inflation and for the equilibrium swollen state. The arrow marks the dry configuration at $P = 1.2487$ from which swelling starts. The swollen equilibrium reached from this state is also computed with \textbf{FEM} and compared below.

\begin{figure}[H]
    \centering                             
    \includegraphics[width=\linewidth]{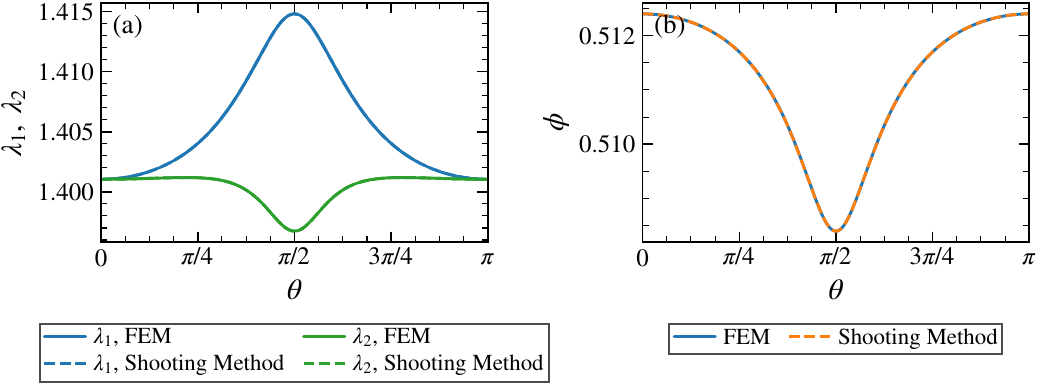}
    \caption{\textbf{Shooting Method} vs.\ \textbf{FEM} at the end of swelling for $P = 1.2487$, $a = 2$, $\mathcal{M} = 0.001$: (a)~deformed shape, $\rho$ vs.\ $\eta$; (b)~meridional stretch $\lambda_1(\theta)$; (c)~circumferential stretch $\lambda_2(\theta)$; (d)~volume fraction $\phi(\theta)$.}
    \label{fig:panels_2x2_method_I_vs_II_M001_a2}
\end{figure}

Fig.~\ref{fig:panels_2x2_method_I_vs_II_M001_a2} compares the final equilibrium states predicted by the two methods. As Fig.~\ref{fig:panels_2x2_method_I_vs_II_M001_a2}(a) shows, the swollen shapes predicted by \textbf{Shooting Method} and \textbf{FEM} are essentially indistinguishable. Figs.~\ref{fig:panels_2x2_method_I_vs_II_M001_a2}(b)--\ref{fig:panels_2x2_method_I_vs_II_M001_a2}(d) give the pointwise comparison along the meridian, plotting $\lambda_1$, $\lambda_2$, and $\phi$ as functions of $\theta$. In each case, the two curves overlap along the entire meridian, so the methods agree both qualitatively and quantitatively.  

This agreement is expected. \textbf{Shooting Method} does not solve the diffusion equation governing the evolution of $\phi$ in $\tau$ and $\theta$, whereas \textbf{FEM} resolves the full chemo-mechanical (diffusion--deformation) coupling. However, both formulations are built on the same free-energy density and therefore share the same equilibrium limit. In both, the strain energy is taken to decompose additively into elastic, mixing, and magnetic contributions, following the approach used for bulk magnetoelastic materials in \cite{garcia2020magneto}; the present results show that this decomposition carries over consistently to the thin-membrane setting. Once swelling is complete, the chemical potential is uniform and the transient coupling no longer affects the final state, so the equilibrium configuration does not depend on the path taken to reach it. The agreement in Fig.~\ref{fig:panels_2x2_method_I_vs_II_M001_a2} therefore provides a mutual verification of the two implementations.

\begin{figure}[H]
    \centering
    \includegraphics[width=\linewidth]{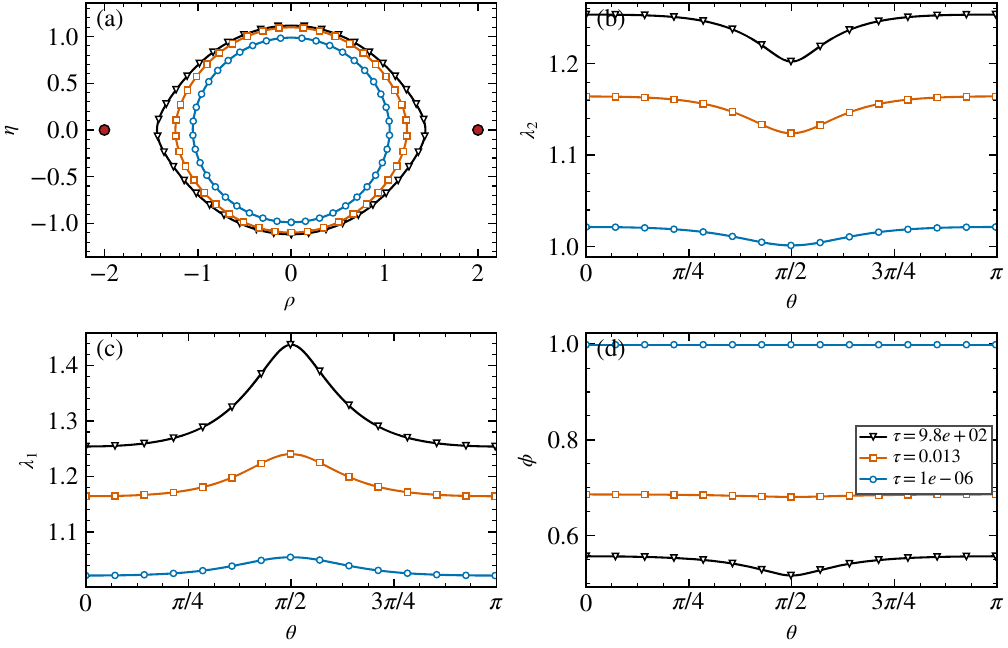}
    \caption{Time evolution of a dry, symmetry-preserving configuration swelling
    at $\mathcal{M} = 0.008$, $a = 2$, and $P = 0.5$: (a)~shape, with the coil
    position marked; (b)~meridional stretch $\lambda_1(\theta)$; (c)~circumferential stretch $\lambda_2(\theta)$; (d)~volume fraction $\phi(\theta)$.}
    \label{fig:panel_2x2_m02_a2_prime}
\end{figure}

\begin{figure}[H]
    \centering
    \includegraphics[width=\linewidth]{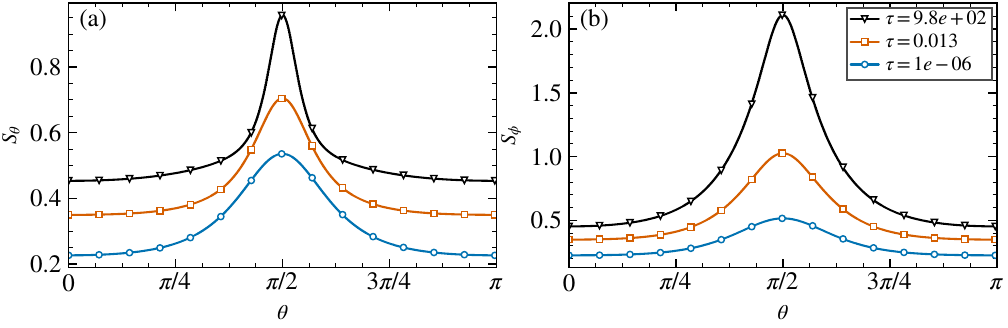}
    \caption{Membrane stresses corresponding to
    $\mathcal{M} = 0.008$, $a = 2$, and $P = 0.5$: (a)~meridional stress $S_\theta(\theta)$; (b)~circumferential stress $S_\phi(\theta)$.}
    \label{fig:S_panels_m02_a2_prime}
\end{figure}

We now turn to the swelling dynamics at the higher field intensity
$\mathcal{M} = 0.008$ and pressure $P = 0.5$, calculated with \textbf{FEM}.
Figure~\ref{fig:panel_2x2_m02_a2_prime}(a) shows the evolution of the balloon
shape, together with the position of the coil. Because the coil lies in the
equatorial plane, the balloon retains the equatorial plane of symmetry of its
initial inflated configuration throughout swelling \citep{shaikh2025inflation}.
As the solvent is absorbed, the balloon grows and becomes increasingly elongated
along the equatorial plane towards the coil. The evolution of $\lambda_2$,
$\lambda_1$, and $\phi$ along the meridian is shown in
Figs.~\ref{fig:panel_2x2_m02_a2_prime}(b)--\ref{fig:panel_2x2_m02_a2_prime}(d). Starting from a nearly unstretched
dry state ($\lambda_1, \lambda_2 = 1.0$--$1.06$, $\phi = 1$), the stretches
increase as the balloon swells. The increase is largest at the equator, where
$\lambda_1$ reaches approximately  $1.44$ at steady state, compared with approximately $1.25$
at the poles. The volume fraction of the polymer decreases from $\phi = 1$ to approximately
$0.56$ at the poles and $0.52$ at the equator, so the equatorial region, which is
closest to the coil, is the most swollen. These results confirm that the
symmetry-preserving dry configuration swells into a symmetry-preserving steady
state. The corresponding membrane stresses $S_\theta$ and $S_\phi$ are shown in
Figs.~\ref{fig:S_panels_m02_a2_prime}(a) and \ref{fig:S_panels_m02_a2_prime}(b). Both stresses increase
throughout swelling and become increasingly concentrated at the equator; at
steady state, $S_\phi$ at the equator ($2.1$) is approximately four to five times
its value at the poles ($0.45$).

\begin{figure}[H]
    \centering
    \includegraphics[width=\linewidth]{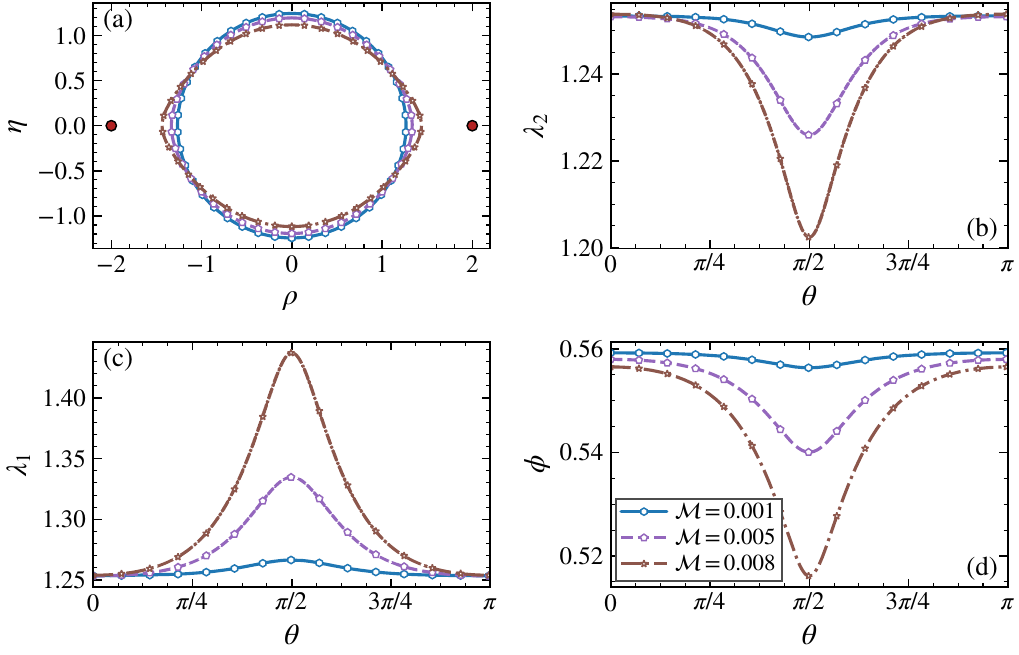}
    \caption{Effect of $\mathcal{M}$ on the final swollen state reached from a
    dry, symmetry-preserving configuration ($a = 2$, $P = 0.5$), for
    $\mathcal{M} = 0.001$, $0.005$, and $0.008$: (a)~shape;
    (b)~circumferential stretch~$\lambda_2(\theta)$; (c)~meridional stretch~$\lambda_1(\theta)$; (d)~polymer volume fraction~$\phi(\theta)$.}
    \label{fig:panel_2x2_M_effect_a2_prime}
\end{figure}

\begin{figure}[H]
    \centering
    \includegraphics[width=\linewidth]{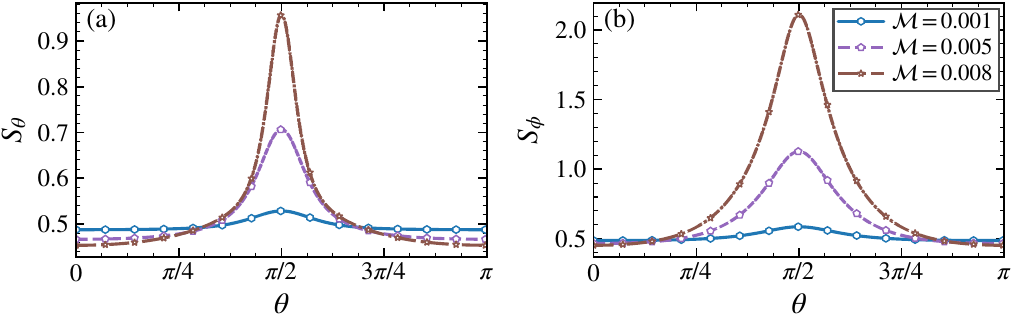}
    \caption{Effect of $\mathcal{M}$ on the final-state membrane stresses
    ($a = 2$, $P = 0.5$), for $\mathcal{M} = 0.001$, $0.005$, and $0.008$:
    (a)~meridional stress $S_\theta(\theta)$; (b)~circumferential stress $S_\phi(\theta)$.}
    \label{fig:S_panels_M_effect_a2_prime}
\end{figure}

Next, we examine how the magnetic field intensity affects the equilibrium
swollen state, starting in each case from a dry, symmetry-preserving inflated
balloon at $P = 0.5$. Figure~\ref{fig:panel_2x2_M_effect_a2_prime}(a) shows the
resulting shapes, and Figs.~\ref{fig:panel_2x2_M_effect_a2_prime}(b)--\ref{fig:panel_2x2_M_effect_a2_prime}(d) show
the corresponding distributions of $\lambda_2$, $\lambda_1$, and $\phi$. The
associated stresses $S_\theta$ and $S_\phi$ are shown in
Figs.~\ref{fig:S_panels_M_effect_a2_prime}(a) and \ref{fig:S_panels_M_effect_a2_prime}(b). As $\mathcal{M}$
increases, the equatorial region is drawn further towards the coil while the
poles move slightly inward. The stretches at the poles are nearly unaffected by
the field ($\lambda_1 = \lambda_2 = 1.25$), whereas at the equator
$\lambda_1$ increases approximately from  $1.27$ to  $1.44$ and $\lambda_2$
decreases from  $1.25$ to  $1.20$ as $\mathcal{M}$ increases from
$0.001$ to $0.008$. The solvent uptake also becomes increasingly non-uniform:
$\phi$ at the equator decreases from  $0.556$ to  $0.516$, while it
remains nearly unchanged at the poles. Correspondingly, the membrane stresses
become strongly concentrated in the equatorial region, where the magnetic
loading is greatest. Over the same range of $\mathcal{M}$, $S_\phi$ at the
equator increases approximately  from  $0.59$ to  $2.1$, while the stresses at
the poles decrease slightly.

\begin{figure}[H]
    \centering
    \includegraphics[width=\linewidth]{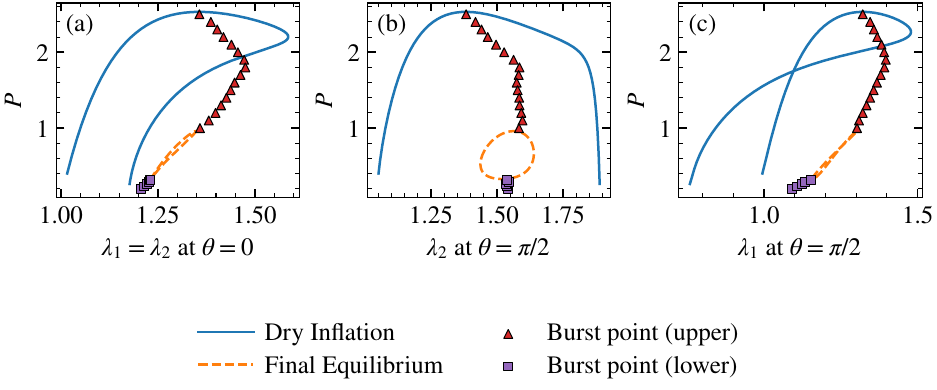}
    \caption{Dry and slow-inflation equilibrium curves and burst points for
    $\mathcal{M} = 0.008$, $a = 2$: (a)~$P$ vs.\ $\lambda_1 = \lambda_2$ at the
    pole ($\theta = 0$); (b)~$P$ vs.\ $\lambda_2$ at the equator
    ($\theta = \pi/2$); (c)~$P$ vs.\ $\lambda_1$ at the equator. Triangles and
    squares denote burst points above and below the slow-inflation isola,
    respectively.}
    \label{fig:Burst_point_a=2}
\end{figure}

Figure~\ref{fig:Burst_point_a=2} summarizes the outcome of swelling at
$\mathcal{M} = 0.008$ over a range of applied pressures, starting in each case
from the dry, symmetry-preserving configuration. The dry and slow-inflation
equilibrium curves are shown in terms of the stretch at the pole Figure~\ref{fig:Burst_point_a=2}(a) and
the stretches at the equator (Figure~\ref{fig:Burst_point_a=2}(b) and \ref{fig:Burst_point_a=2}(c)). At this field intensity, the
slow-inflation curve forms a closed loop (an isola) spanning $0.33 \leq P \leq 0.96$.
For pressures within this range, the balloon swells into a stable equilibrium
configuration on the closed loop (isola), as in Fig.~\ref{fig:panel_2x2_m02_a2_prime}. Outside
this range, no swollen equilibrium is available, and the balloon bursts. The
burst points therefore form two families: an upper family for
$1.0 \leq P \leq 2.5$, above the closed loop (isola), and a lower family for
$0.2 \leq P \leq 0.32$, below it. Each family terminates close to the
corresponding extremity of the closed loop.

The two families show opposite trends with pressure. For the upper family, the
time to burst increases as the pressure decreases towards the closed loop (isola) (not shown).
The stretch at burst also increases as $P$ decreases from $2.5$ to about $1.9$,
reaching $\lambda_2 = 1.48$ at the pole. At lower pressures, the stretch at
the pole and $\lambda_1$ at the equator decrease again towards the upper
extremity of the closed loop (isola), whereas $\lambda_2$ at the equator remains nearly
constant at  $1.58$. For the lower family, both the time to burst and the
stretch at burst increase with pressure: as $P$ increases from $0.2$ to $0.32$,
the stretch at the pole at burst increases approximately from  $1.21$ to
 $1.23$, and $\lambda_1$ at the equator from  $1.09$ to
 $1.16$, while $\lambda_2$ at the equator remains at $1.54$.

The magnetic field thus controls the range of pressure over which a stable
swollen configuration exists. For $\mathcal{M} \geq 0.006$, the
slow-inflation curve becomes a closed loop (an isola), with pressure upper and lower bound approximately $0.11 \leq P \leq 1.22$ for $\mathcal{M} = 0.006$,
$0.33 \leq P \leq 0.96$ for $\mathcal{M} = 0.008$, and
$0.56 \leq P \leq 0.71$ for $\mathcal{M} = 0.0088$. Increasing the
magnetic field intensity therefore shrinks the pressure range of stable swollen
configurations from both above and below, so that, for $a = 2$, swollen
equilibria exist only within an increasingly narrow band of pressure.

\subsubsection{Symmetry-breaking magnetic time evolution}
\label{sec: Symmetry-breaking magnetic time evolution}

\begin{figure}[!ht]
    \centering                         
    \includegraphics[width=\linewidth]{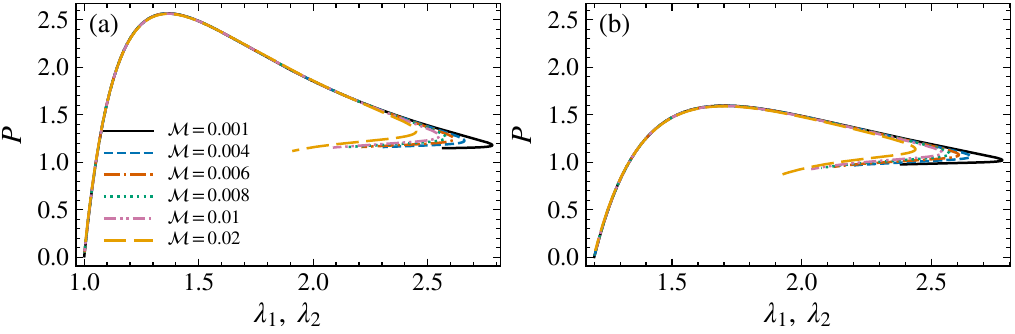}
    \caption{Pressure--stretch curves at the pole ($P$ vs.\ $\lambda_1$, $\lambda_2$) for $a = 3$ and various values of $\mathcal{M}$: (a)~dry inflation; (b)~final equilibrium configuration.}
    \label{fig:P_vs_lambda2_a3_pole}
\end{figure}

We now repeat the analysis with the coil radius increased to $a = 3$. The solvent parameter is again $\nu = 0.10668$ ($\lambda_0 = 1.2$).
Fig.~\ref{fig:P_vs_lambda2_a3_pole} shows the pressure--stretch curves at the pole, $P$ vs.\ $\lambda_1$ and $\lambda_2$, for dry inflation(Fig.~\ref{fig:P_vs_lambda2_a3_pole}(a)) and final equilibrium configuration (Fig.~\ref{fig:P_vs_lambda2_a3_pole}(b)), i.e.\ the final swollen configuration. for several values of $\mathcal{M}$. 

\begin{figure}[H]
    \centering
    \includegraphics[width=0.6\linewidth]{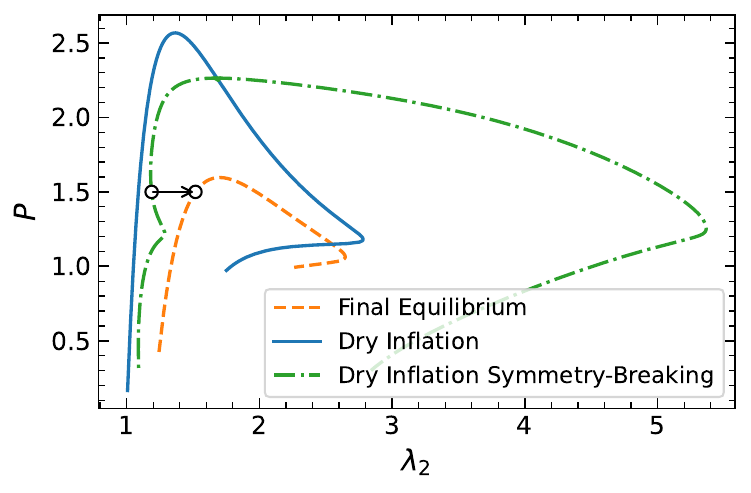}
    \caption{Pressure--stretch curves ($P$ vs.\ $\lambda_2$ at the pole) for $\mathcal{M} = 0.001$, $a = 3$: dry inflation (solid blue) and equilibrium swelling via \textbf{Shooting Method} (dashed orange). The arrow marks the swelling process starting from the dry symmetry-breaking configuration at $P = 1.5$.}
    \label{fig:P_vs_lambda2_SB_a3}
\end{figure}

At $a = 3$ and $\mathcal{M} = 0.001$, \textbf{Shooting Method} (\ref{sec:Appendix}) returns \emph{two} dry-inflation equilibrium branches: one symmetry-preserving and one symmetry-breaking. We therefore consider an inflated, dry balloon on the symmetry-breaking branch, held at constant pressure $P = 1.5$.
Fig.~\ref{fig:P_vs_lambda2_SB_a3} shows the pole pressure--stretch curves at $\mathcal{M} = 0.001$ computed with \textbf{Shooting Method}, for dry inflation and for the equilibrium swollen state. The arrow marks the dry symmetry-breaking configuration from which swelling starts.

Both methods start from this same symmetry-breaking configuration, but reach the swollen state, which is symmetric in nature. \textbf{FEM} (\ref{Method_II}) considers the coupled diffusion--deformation problem and integrates forward in time, following the physical swelling path as the gel absorbs solvent. The balloon does not stay on the symmetry-breaking branch. Symmetry is recovered during the transient, and the solution converges to the symmetric slow-inflation curve at $P = 1.5$, where the chemical potential of the gel equilibrates with that of the surrounding solvent. \textbf{Shooting Method} instead determines equilibrium states directly by energy minimisation: it does not solve the diffusion equation, and time does not enter the formulation. Using the same configuration as the initial iterate for the swollen equilibrium problem, it converges to the same symmetric state.

\begin{figure}[H]
    \centering                            
    \includegraphics[width=0.85\linewidth]{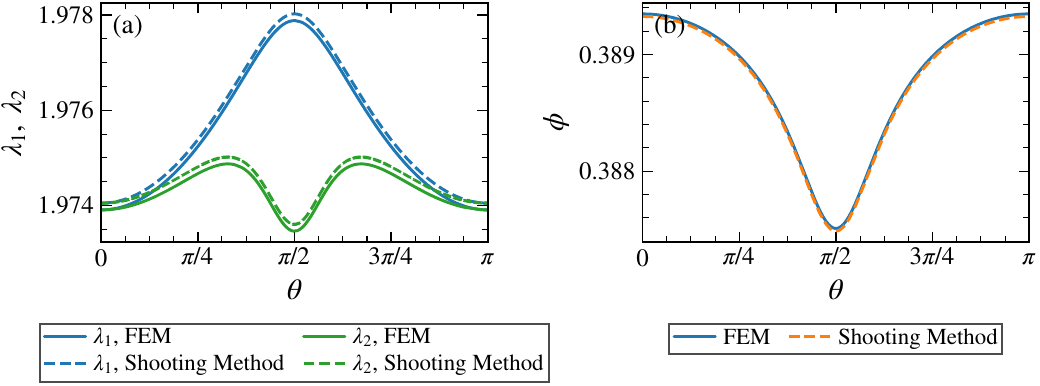}
    \caption{\textbf{Shooting Method} vs.\ \textbf{FEM} at the end of swelling for $P = 1.285$, $a = 3$, $\mathcal{M} = 0.001$, both started from the same inflated symmetry-breaking configuration at $P = 1.5$: (a)~shape, $\rho$ vs.\ $\eta$; (b)~meridional stretch~$\lambda_1(\theta)$; (c)~circumferential stretch~$\lambda_2(\theta)$; (d)~polymer volume fraction~$\phi(\theta)$.}
    \label{fig:panels_2x2_method_I_vs_II_M001_a3_SB}
\end{figure}

The final swollen shapes coincide in Fig.~\ref{fig:panels_2x2_method_I_vs_II_M001_a3_SB}(a), as do the meridional profiles of $\lambda_1$, $\lambda_2$, and $\phi$ in Figs.~\ref{fig:panels_2x2_method_I_vs_II_M001_a3_SB}(b)--\ref{fig:panels_2x2_method_I_vs_II_M001_a3_SB}(d). This agreement is a mutual verification of the two formulations: one reaches the swollen equilibrium by minimizing the total energy, the other by integrating the transient diffusion--deformation problem. Both converge to the same symmetric configuration despite starting from a symmetry-breaking one, which indicates that, at these parameters, the symmetry-breaking state does not persist through swelling. 

\begin{figure}[H]
  \centering                           
  \includegraphics[width=\linewidth]{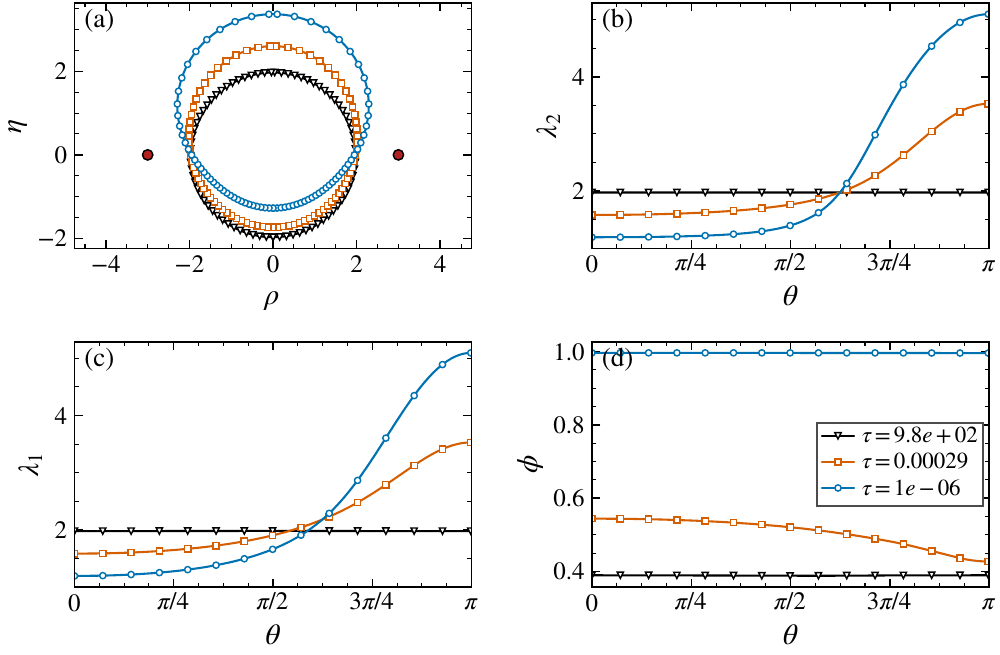}
  \caption{Time evolution from a dry symmetry-breaking configuration at $\mathcal{M} = 0.015$, $a = 3$, $\lambda_0 = 1.2$: (a)~shape, $\rho$ vs.\ $\eta$, with the coil at $(\pm a, 0)$; (b)~circumferential stretch~$\lambda_2(\theta)$; (c)~meridional stretch~$\lambda_1(\theta)$; (d)polymer volume fraction~$\phi(\theta)$.}
  \label{fig:panel_2x2_m02_a3_sb}
\end{figure}

\begin{figure}[H]
  \centering                       
  \includegraphics[width=\linewidth]{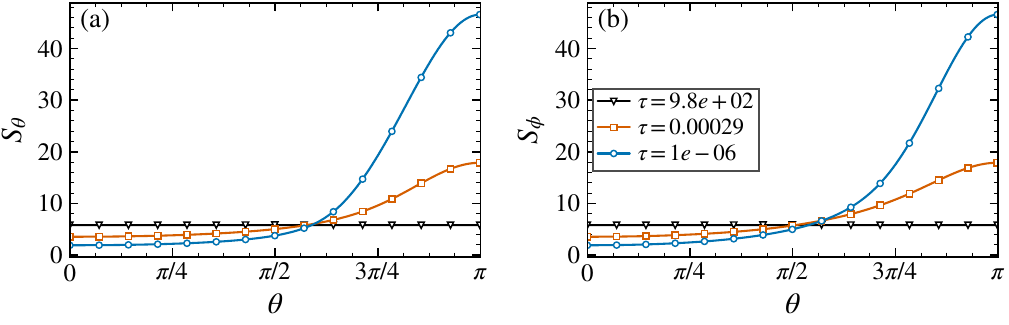}
  \caption{Stress evolution corresponding to Fig.~\ref{fig:panel_2x2_m02_a3_sb}, shown at three snapshots: (a)~meridional stress $S_\theta(\theta)$; (b)~circumferential stress $S_\phi(\theta)$.}
  \label{fig:S_panels_m02_a3_sb}
\end{figure}

We next consider the swelling dynamics at a higher field intensity, $\mathcal{M} = 0.015$, and $P = 1.5$, again computed with \textbf{FEM}. 
Fig.~\ref{fig:panel_2x2_m02_a3_sb}(a) shows the time evolution of the balloon shape, with the coil cross-section drawn to scale, and Figs.~\ref{fig:panel_2x2_m02_a3_sb}(b)--\ref{fig:panel_2x2_m02_a3_sb}(d) show the evolution of $\lambda_2$, $\lambda_1$, and $\phi$, respectively. In this case, the dry symmetry-breaking configuration swells into a symmetric swollen state. The associated membrane stresses $S_\theta$ and $S_\phi$ are shown in Figs.~\ref{fig:S_panels_m02_a3_sb}(a) and \ref{fig:S_panels_m02_a3_sb}(b). Here \textbf{FEM}  provides what \textbf{Shooting Method} cannot: the transient through which symmetry is recovered and the time scale over which this occurs.

\begin{figure}[H]
  \centering                         
  \includegraphics[width=\linewidth]{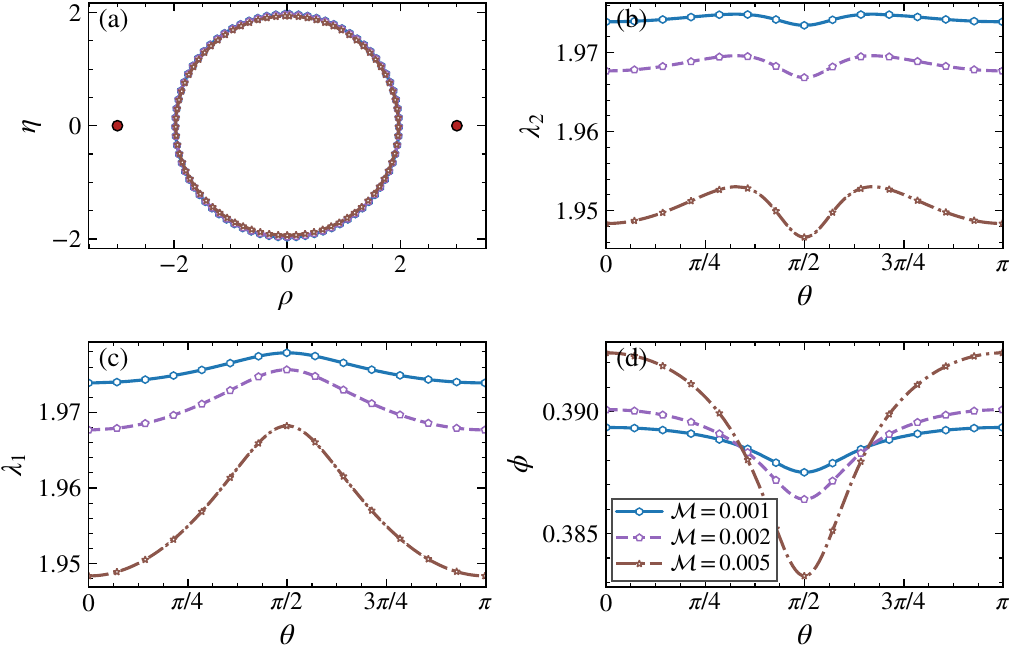}
  \caption{Effect of $\mathcal{M}$ on the equilibrium state reached from a dry symmetry-breaking configuration ($a = 3$, $\lambda_0 = 1.2$), for $\mathcal{M} = 0.001$, $0.01$, and $0.015$: (a)~shape, with the coil at $(\pm a, 0)$; (b)~circumferential stretch~$\lambda_2(\theta)$; (c)~meridional stretch~$\lambda_1(\theta)$; (d)polymer volume fraction~$\phi(\theta)$.}
  \label{fig:panel_2x2_M_effect_a3_sb}
\end{figure}

\begin{figure}[H]
  \centering                             
  \includegraphics[width=0.95\linewidth]{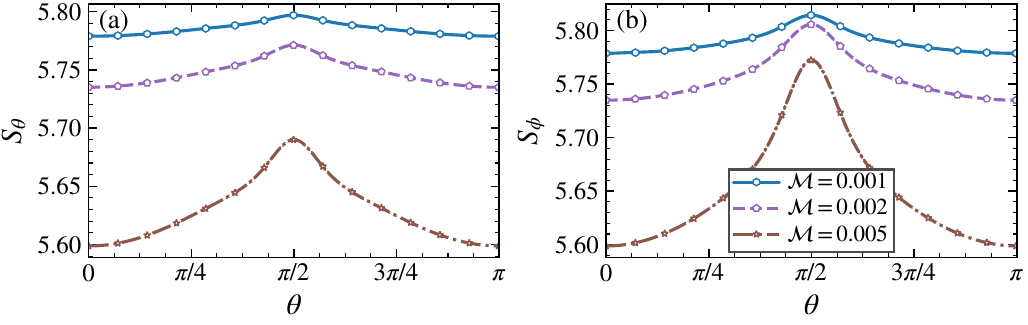}
  \caption{Effect of $\mathcal{M}$ on the equilibrium stresses ($a = 3$, $\lambda_0 = 1.2$), for $\mathcal{M} = 0.001$, $0.01$, and $0.015$: (a)~meridional stress $S_\theta(\theta)$; (b)~circumferential stress $S_\phi(\theta)$.}
  \label{fig:S_panels_M_effect_a3_sb}
\end{figure}

Finally, we examine how the magnetic field intensity affects the equilibrium swollen state, starting in each case from the dry symmetry-breaking inflated configuration. Fig.~\ref{fig:panel_2x2_M_effect_a3_sb} shows the resulting balloon shapes and the profiles of $\lambda_2$, $\lambda_1$, and $\phi$, and Fig.~\ref{fig:S_panels_M_effect_a3_sb} shows the corresponding stresses $S_\theta$ and $S_\phi$. As in the symmetric case, increasing $\mathcal{M}$ pulls the equilibrium configuration towards the coil and sharpens the meridional gradients in stretch and solvent fraction. In this case, however, the symmetry-breaking character is retained at all field strengths considered, and it amplifies the contrast between the poles ($\theta = 0, \pi$) and the equator ($\theta = \pi/2$): the redistribution of $S_\theta$ and $S_\phi$ between these regions becomes more pronounced as $\mathcal{M}$ increases.

\begin{figure}[H]
    \centering                         
    \includegraphics[width=\linewidth]{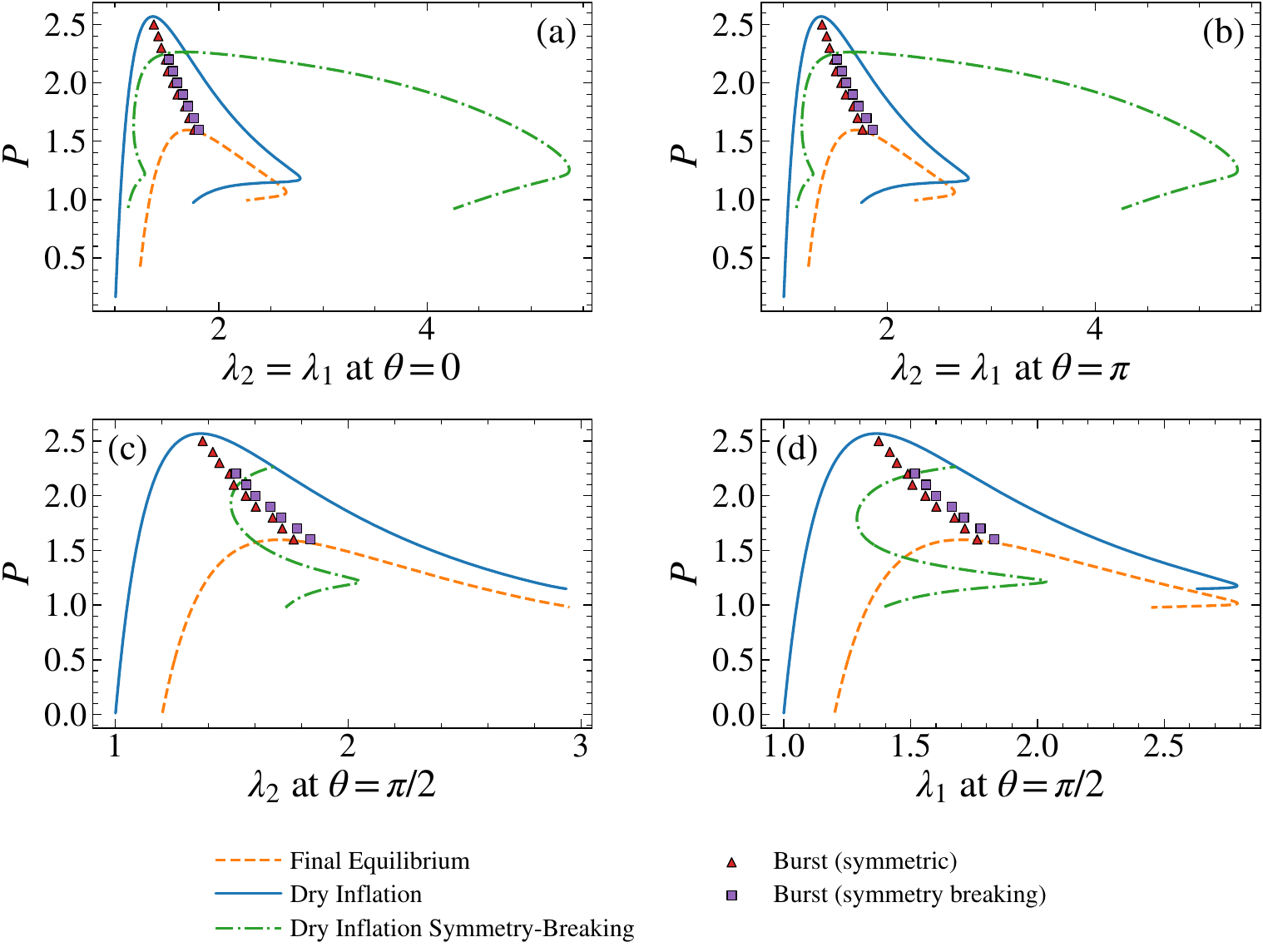}
    \caption{Dry (symmetric and symmetry-breaking) and slow-inflation
    equilibrium curves with burst points for $\mathcal{M} = 0.001$, $a = 3$:
    (a)~$P$ vs.\ $\lambda_1 = \lambda_2$ at $\theta = 0$;
    (b)~$P$ vs.\ $\lambda_1 = \lambda_2$ at $\theta = \pi$;
    (c)~$P$ vs.\ $\lambda_2$ at $\theta = \pi/2$;
    (d)~$P$ vs.\ $\lambda_1$ at $\theta = \pi/2$. Triangles and squares denote
    burst points for swelling from dry symmetric and symmetry-breaking
    configurations, respectively.}
    \label{fig:Burst_point_a=3}
\end{figure}

Figure~\ref{fig:Burst_point_a=3} shows the corresponding results for a coil of
radius $a = 3$ at $\mathcal{M} = 0.001$. Panels (a) and (b) show the dry
symmetric, dry symmetry-breaking, and slow-inflation equilibrium curves in terms
of the stretch at the poles $\theta = 0$ and $\theta = \pi$. Panels (c) and (d)
show the same curves in terms of the stretches $\lambda_2$ and $\lambda_1$ at
the equator. Burst points are shown for swelling that starts from both dry
symmetric (triangles) and dry symmetry-breaking (squares) configurations. In
contrast to the case $a = 2$, $\mathcal{M} = 0.008$, the slow-inflation curve
here is not an isola. Its limit-point pressure is approximately  $P = 1.60$, and burst
occurs only above this pressure, so no lower family of burst points appears.
Burst points starting from symmetric configurations span
$1.6 \leq P \leq 2.5$. Those starting from symmetry-breaking configurations span
$1.6 \leq P \leq 2.2$, because the dry symmetry-breaking branch exists only
below its maximum pressure, appropriately $P = 2.26$.

For both families, the stretch at burst increases monotonically as the pressure
decreases towards the limit-point pressure of the slow-inflation curve. For
swelling from symmetric configurations, the stretches at the poles and at the
equator are nearly equal at burst. They increase from $\lambda_2 =  1.37$ at
$P = 2.5$ to $\lambda_2 =  1.76$ at $P = 1.6$, indicating that the balloon reaches
almost spgerical just before burst. For swelling from symmetry-breaking
configurations, the stretches at burst are larger than in the symmetric case at
the same pressure and are slightly non-uniform. At $P = 1.6$, for example, they
are approximately  $\lambda_2 =  1.81$ at $\theta = 0$, $\lambda_2 = 1.86$ at $\theta = \pi$, and
$\lambda_2 = 1.83$ at the equator, compared with $\lambda_2 = 1.76$ for the symmetric
case. This difference decreases as the pressure increases: at $P = 2.2$, the
stretches at burst are approximately  $\lambda_2 = 1.52$ for the symmetry-breaking case and
$\lambda_2 =  1.49$ for the symmetric case. The time to burst also increases as the
pressure decreases (not shown). For this weak field, the burst response
therefore closely resembles that of the non-magnetic balloon
(Fig.~\ref{fig:Equator_pole1}).

\subsection{Two-coil arrangement}
\label{sec:two_coil}

We now load the balloon with two coaxial coils placed at heights $h = \pm h_a$, with $h_a = \sqrt{2}$, above and below the equatorial plane. The radii and field intensities of the upper and lower coils are denoted by $a_\mathrm{u}, a_\mathrm{l}$ and $\mathcal{M}_\mathrm{u}, \mathcal{M}_\mathrm{l}$, respectively. We consider three cases: (i)~identical coils carrying equal currents, which load the balloon symmetrically about the equator; (ii)~equal radii but unequal currents; and (iii)~equal currents but unequal radii. In cases (ii) and (iii) the asymmetry is imposed by the loading. This distinguishes them from the spontaneous symmetry breaking of section~\ref{sec: Symmetry-breaking magnetic time evolution}.

The results for each case are presented in the same way. We first give the pressure--stretch curves for dry inflation configuration and final equilibrium configuration at the poles and at the equator. We then give the time evolution of swelling at constant pressure, computed with \textbf{FEM} with $\lambda_0 = 1.2$. 

\subsubsection{Symmetric magnetic loading}
\label{sec:two_coil_symmetric}

\begin{figure}[H]
    \centering                            
    \includegraphics[width=\linewidth]{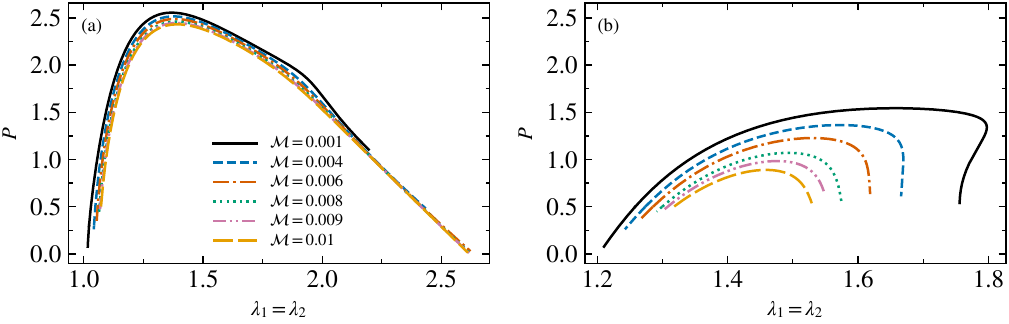}
    \caption{Pressure--stretch curves ($P$ vs.\ $\lambda_1$, $\lambda_2$) at the pole under symmetric two-coil loading ($a_\mathrm{u} = a_\mathrm{l} = \sqrt{2}$, $\mathcal{M}_\mathrm{u} = \mathcal{M}_\mathrm{l} = \mathcal{M}$), for various $\mathcal{M}$: (a)~dry inflation; (b)~final equilibrium configuration.}
    \label{fig:P_vs_lambda_pole_Dry_slow_TC}
\end{figure}

Both coils have radius $a_\mathrm{u} = a_\mathrm{l} = \sqrt{2}$ and equal intensity $\mathcal{M}_\mathrm{u} = \mathcal{M}_\mathrm{l} = \mathcal{M}$. Since the loading is symmetric about the equator, the two poles respond identically. Fig.~\ref{fig:P_vs_lambda_pole_Dry_slow_TC} shows the pressure--stretch curves at the pole, for several values of $\mathcal{M}$, for dry inflation(Fig.~\ref{fig:P_vs_lambda_pole_Dry_slow_TC} (a)) and for the final equilibrium configuration(Fig.~\ref{fig:P_vs_lambda_pole_Dry_slow_TC} (b)). It is seen that the effect of increasing the magnetic field is significant only at larger stretches in the case of the dry inflation configuration(Fig.~\ref{fig:P_vs_lambda_pole_Dry_slow_TC} (a)). The Increase in magnetic field intensity decreases limit point pressure for dry as well as final equilibrium configuration, but reduction in limit point pressure in the case of final equilibrium configuration(Fig.~\ref{fig:P_vs_lambda_pole_Dry_slow_TC} (b)) is much more significant compared to dry inflation configuration.     

\begin{figure}[H]
  \centering
  \includegraphics[width=\linewidth]{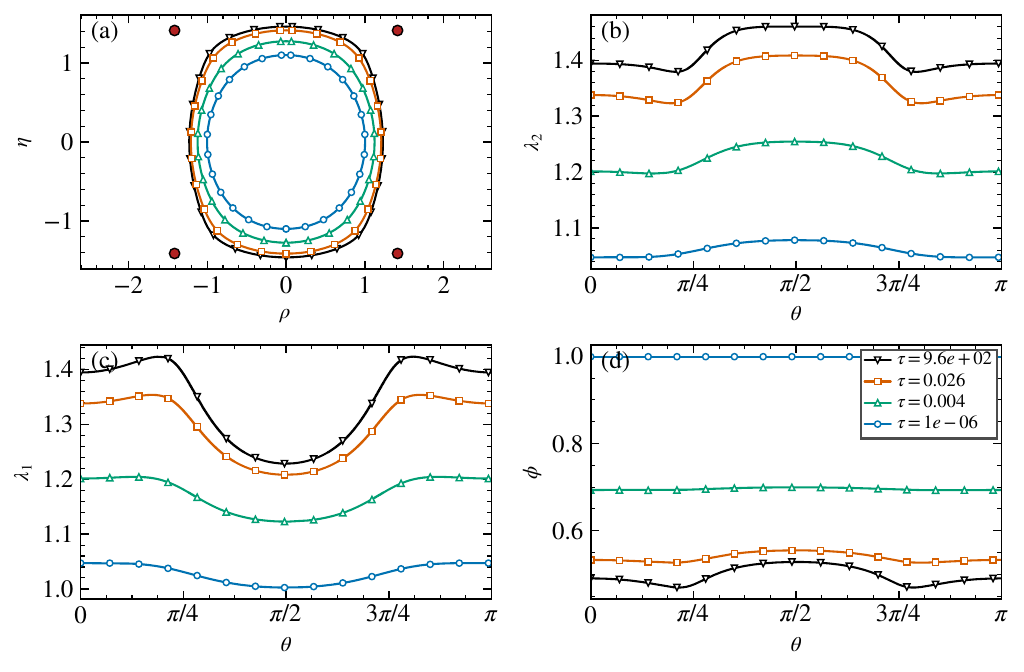}
  \caption{Swelling from the dry state under symmetric two-coil loading ($a = \sqrt{2}$, $\mathcal{M} = 0.01$, $P = 0.8$): (a)~shape, with the coils at $(\pm a, \pm h)$; (b)~circumferential stretch~$\lambda_2(\theta)$; (c)~meridional stretch~$\lambda_1(\theta)$; (d)~polymer volume fraction~$\phi(\theta)$.}
  \label{fig:panel_2x2_m01_TC_equal}
\end{figure}

\begin{figure}[H]
    \centering                         
    \includegraphics[width=\linewidth]{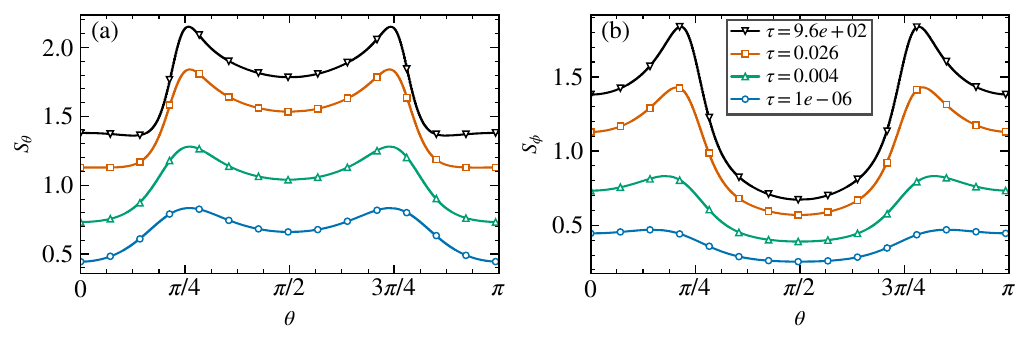}
    \caption{Membrane stresses for the case of Fig.~\ref{fig:panel_2x2_m01_TC_equal}: (a)~meridional stress $S_\theta(\theta)$; (b)~circumferential stress $S_\phi(\theta)$.}
    \label{fig:S_panels_m01_TC_equal}
\end{figure}

Figs.~\ref{fig:panel_2x2_m01_TC_equal} and~\ref{fig:S_panels_m01_TC_equal} show the time evolution of swelling at $\mathcal{M} = 0.01$ and $P = 0.8$. Each coil pulls the part of the membrane nearest to it, so the balloon deforms locally at two sites. This is the attraction seen in the single-coil case, now acting at two locations. The stresses $S_\theta$ and $S_\phi$ peak correspondingly in the regions closest to the coils. Because the loading is symmetric, the shape, stretches, solvent fraction, and stresses all remain symmetric about the equator throughout swelling.

\begin{figure}[H]
    \centering
    \includegraphics[width=\linewidth]{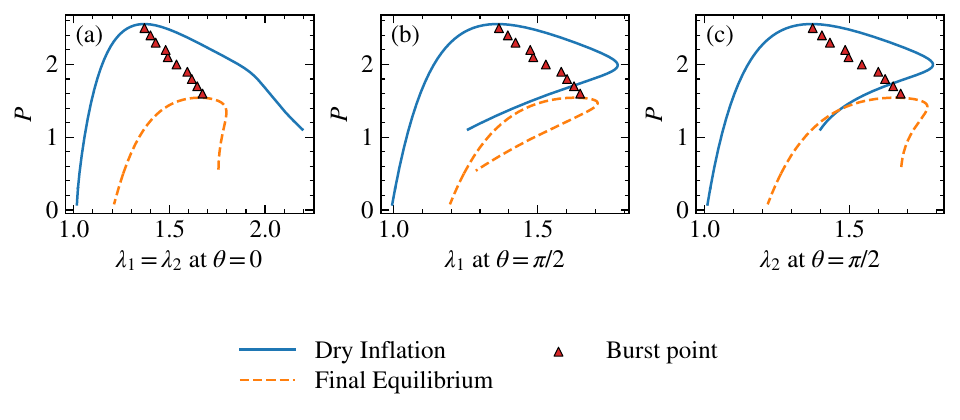}
    \caption{Dry and slow-inflation equilibrium curves with burst points under
    symmetric two-coil loading ($a_\mathrm{u} = a_\mathrm{l} = \sqrt{2}$,
    $\mathcal{M}_\mathrm{u} = \mathcal{M}_\mathrm{l} = 0.001$):
    (a)~$P$ vs.\ $\lambda_1 = \lambda_2$ at the pole ($\theta = 0$);
    (b)~$P$ vs.\ $\lambda_1$ at the equator ($\theta = \pi/2$);
    (c)~$P$ vs.\ $\lambda_2$ at the equator. Triangles denote burst points.}
    \label{fig:Burst_point_equal}
\end{figure}

Figure~\ref{fig:Burst_point_equal} shows the dry and slow-inflation
equilibrium curves together with the burst points for symmetric two-coil
loading at $\mathcal{M} = 0.001$. Because the loading is symmetric about the
equator, the poles $\theta = 0$ and $\theta = \pi$ respond identically, so
only the pole $\theta = 0$ is shown in Figure~\ref{fig:Burst_point_equal}(a). Figures~\ref{fig:Burst_point_equal}(b) and  ~\ref{fig:Burst_point_equal}(c) show
the stretches $\lambda_1$ and $\lambda_2$ at the equator. The dry curve
reaches a maximum pressure of $P = 2.56$, and the slow-inflation
curve has its limit point at $P = 1.55$. The burst points occurs for pressure values
spanning $1.6 \leq P \leq 2.5$.

The stretch at burst increases monotonically as the pressure decreases
towards the limit point of the slow-inflation curve. At the pole, it
increases from $\lambda_2 =  1.37$ at $P = 2.5$ to $\lambda_2  = 1.67$ at $P = 1.6$.
At the equator, $\lambda_1$ increases from  $1.37$ to $1.65$,
and $\lambda_2$ from $ 1.37$ to $1.68$, over the same range.
The stretches at the pole and at the equator are therefore nearly equal at
burst. For this weak
field, the burst response of the symmetric two-coil arrangement thus closely
resembles that of the non-magnetic balloon (Fig.~\ref{fig:Equator_pole1}).

\subsubsection{Asymmetric loading: unequal field intensities}
\label{sec:two_coil_asymmetric}

\begin{figure}[H]
    \centering
    \includegraphics[width=\linewidth]{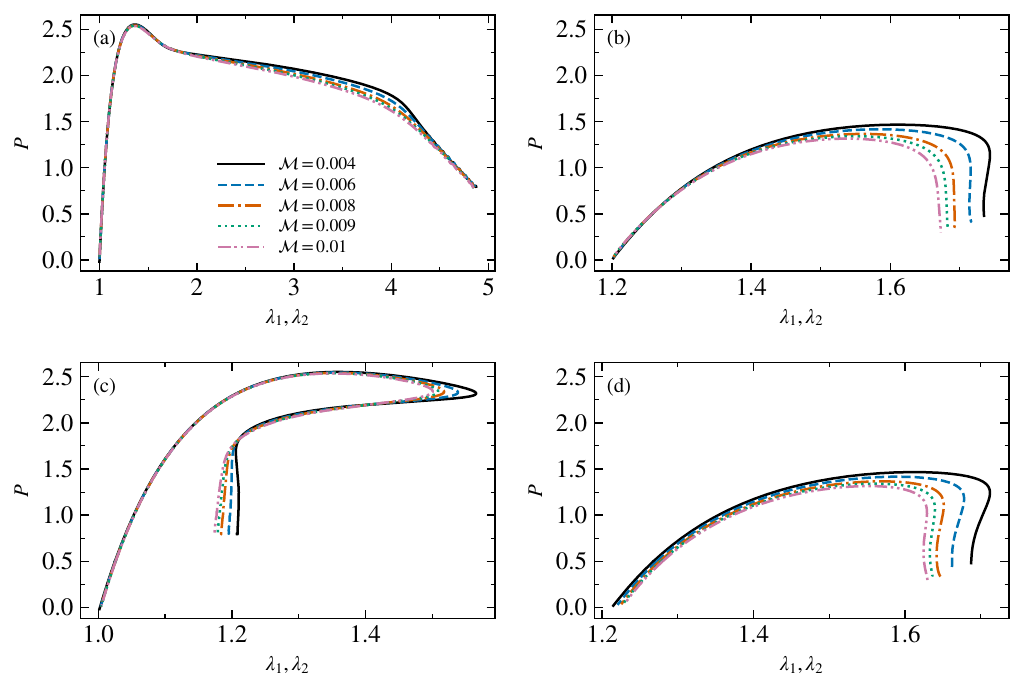}
    \caption{Pressure--stretch curves ($P$ vs.\ $\lambda_1$, $\lambda_2$) at the poles for unequal intensities ($\mathcal{M}_\mathrm{l} = 0.001$, various $\mathcal{M}_\mathrm{u}$): (a),~(b)~pole $\theta = 0$, dry inflation equilibrium and final equilibrium configuration; (c),~(d)~pole $\theta = \pi$, dry inflation equilibrium and final equilibrium configuration.}
    \label{fig:P_vs_lambda_poles_TC_Diff_M}
\end{figure}

The coils keep equal radii, $a_\mathrm{u} = a_\mathrm{l} = \sqrt{2}$, but carry different currents. The lower-coil intensity is fixed at $\mathcal{M}_\mathrm{l} = 0.001$, and $\mathcal{M}_\mathrm{u}$ is varied. The two poles now respond differently, so Fig.~\ref{fig:P_vs_lambda_poles_TC_Diff_M} shows the pressure--stretch curves at both $\theta = 0$ and $\theta = \pi$.

\begin{figure}[H]
  \centering                            
  \includegraphics[width=\linewidth]{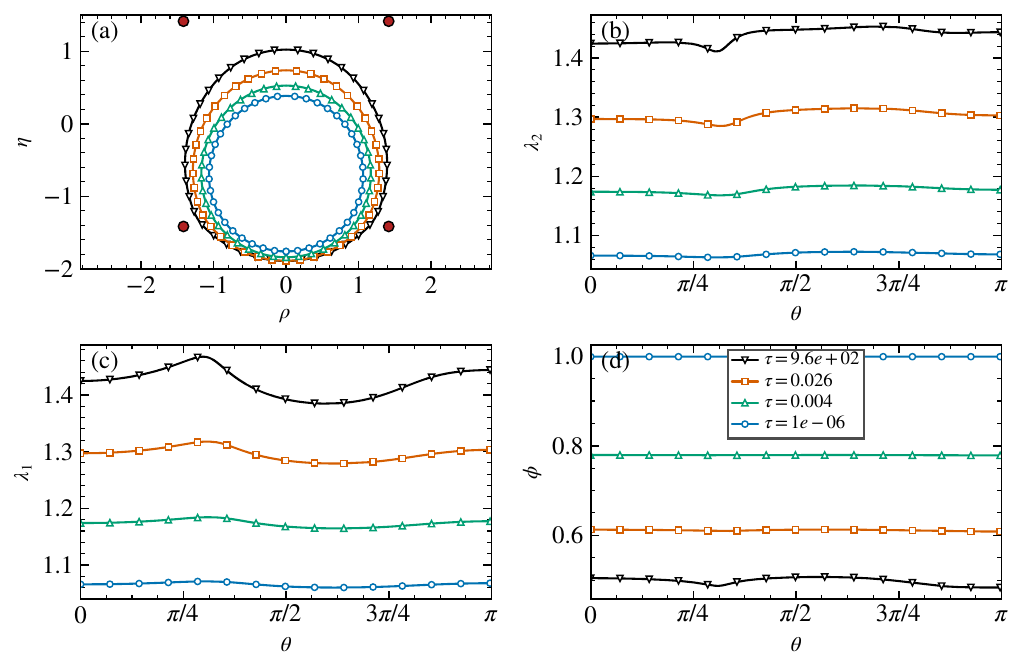}
  \caption{Swelling from the dry state under unequal intensities ($a = \sqrt{2}$, $\mathcal{M}_\mathrm{u} = 0.01$, $\mathcal{M}_\mathrm{l} = 0.001$, $P = 1.2$); (a)~shape; (b)~circumferential stretch~$\lambda_2(\theta)$; (c)~meridional stretch~$\lambda_1(\theta)$; (d)~polymer volume fraction~$\phi(\theta)$.}
  \label{fig:panel_2x2_m01_TC_diff_M}
\end{figure}

\begin{figure}[H]
    \centering
    \includegraphics[width=\linewidth]{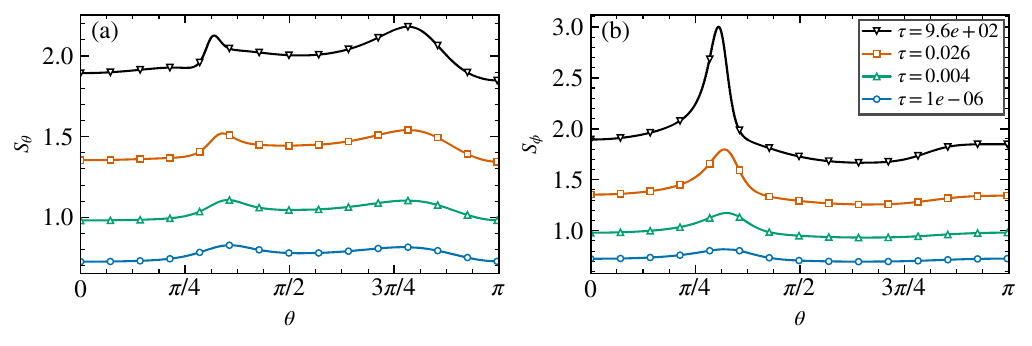}
    \caption{Membrane stresses for the unequal intensities ($a = \sqrt{2}$, $\mathcal{M}_\mathrm{u} = 0.01$, $\mathcal{M}_\mathrm{l} = 0.001$, $P = 1.2$): (a)~meridional stress $S_\theta(\theta)$; (b)~circumferential stress $S_\phi(\theta)$.}
    \label{fig:S_panels_m01_TC_diff_M}
\end{figure}

Figs.~\ref{fig:panel_2x2_m01_TC_diff_M} and~\ref{fig:S_panels_m01_TC_diff_M} show the time evolution of swelling at $\mathcal{M}_\mathrm{u} = 0.01$ and $P = 1.2$. As in the symmetric case, each coil attracts the nearest part of the membrane. The pull towards the upper coil, which carries the larger current, is now much stronger, so the two local deformations differ markedly, and the dominant stress peaks occur near the upper coil. Although the coils are geometrically identical and symmetrically placed, their unequal currents make the loading asymmetric. As a result, the shape, stretches, solvent fraction, and stresses remain asymmetric about the equator throughout swelling.

\begin{figure}[H]
    \centering
    \includegraphics[width=0.8\linewidth]{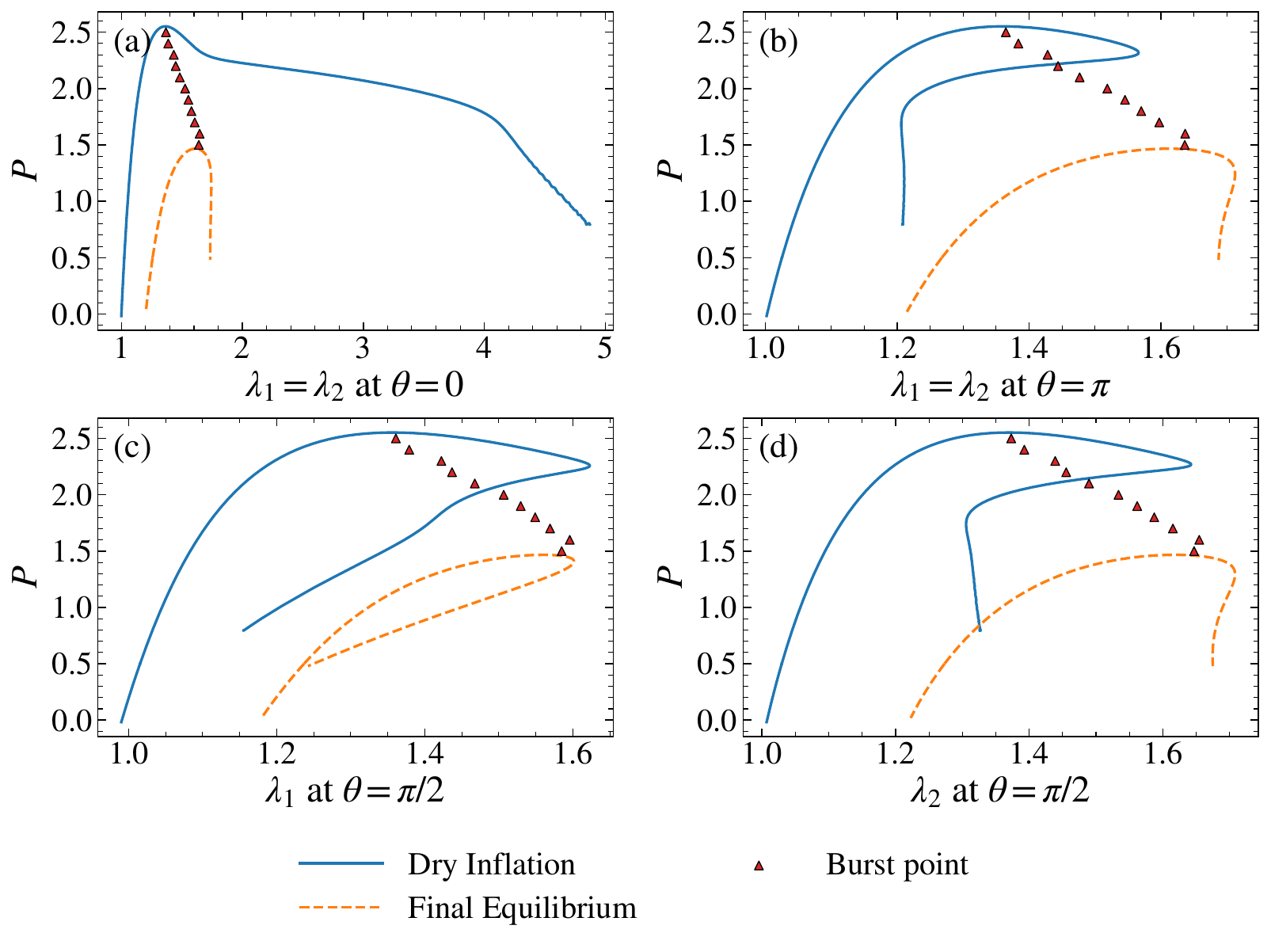}
    \caption{Dry and slow-inflation equilibrium curves with burst points for
    unequal intensities ($a_\mathrm{u} = a_\mathrm{l} = \sqrt{2}$,
    $\mathcal{M}_\mathrm{u} = 0.004$, $\mathcal{M}_\mathrm{l} = 0.001$):
    (a)~$P$ vs.\ $\lambda_1 = \lambda_2$ at $\theta = 0$;
    (b)~$P$ vs.\ $\lambda_1 = \lambda_2$ at $\theta = \pi$;
    (c)~$P$ vs.\ $\lambda_1$ at $\theta = \pi/2$;
    (d)~$P$ vs.\ $\lambda_2$ at $\theta = \pi/2$. Triangles denote burst
    points.}
    \label{fig:Burst_point_M_unequal}
\end{figure}

Figure~\ref{fig:Burst_point_M_unequal} shows the dry and slow-inflation
equilibrium curves with burst points for unequal intensities,
$\mathcal{M}_\mathrm{u} = 0.004$ and $\mathcal{M}_\mathrm{l} = 0.001$.
Figs.~\ref{fig:Burst_point_M_unequal}(a) and  ~\ref{fig:Burst_point_M_unequal}(b) show the stretch at the poles $\theta = 0$ and
$\theta = \pi$, and Figs.~\ref{fig:Burst_point_M_unequal}(c) and ~\ref{fig:Burst_point_M_unequal}(d) show $\lambda_1$ and $\lambda_2$ at the equator. The asymmetry of the loading is most visible on the dry curve:
beyond limit point pressure, the stretch at $\theta = 0$
grows to $\lambda_1 = \lambda_2 =  4.9$ (Fig.~\ref{fig:P_vs_lambda_poles_TC_Diff_M}(a)) , whereas the stretch at $\theta = \pi$
remains below $\lambda_1 = \lambda_2 = 1.6$ (Fig.~\ref{fig:P_vs_lambda_poles_TC_Diff_M}(c)). The slow-inflation curve has its limit point approximately  at $P = 1.47$ (Fig.~\ref{fig:P_vs_lambda_poles_TC_Diff_M}(b)), so pressure above this value delayed burst occurs, and the burst points span $1.5 \leq P \leq 2.5$.

In contrast to the dry curve, the burst points show only a weak asymmetry.
The stretch at burst increases as the pressure decreases towards the limit
point of the slow-inflation curve. At $\theta = 0$, it increases from
$\lambda_2 = 1.37$ at $P = 2.5$ to $\lambda_2 =  1.64$ at $P = 1.5$, and at
$\theta = \pi$ from $\lambda_2 =  1.36$ to $\lambda_2 = 1.63$. At the equator,
$\lambda_2$ increases approximately from  $1.37$ to $1.65$, while
$\lambda_1$ increases from $1.36$ to $1.60$ at $P = 1.6$ and
then decreases slightly to $1.58$ at $P = 1.5$. The stretches at the
two poles and at the equator therefore remain close to one another at burst,
even though the loading is asymmetric.

\subsubsection{Asymmetric loading: unequal coil radii}
\label{sec:two_coil_asymmetric_radius}

\begin{figure}[htbp]
    \centering
    \includegraphics[width=0.9\linewidth]{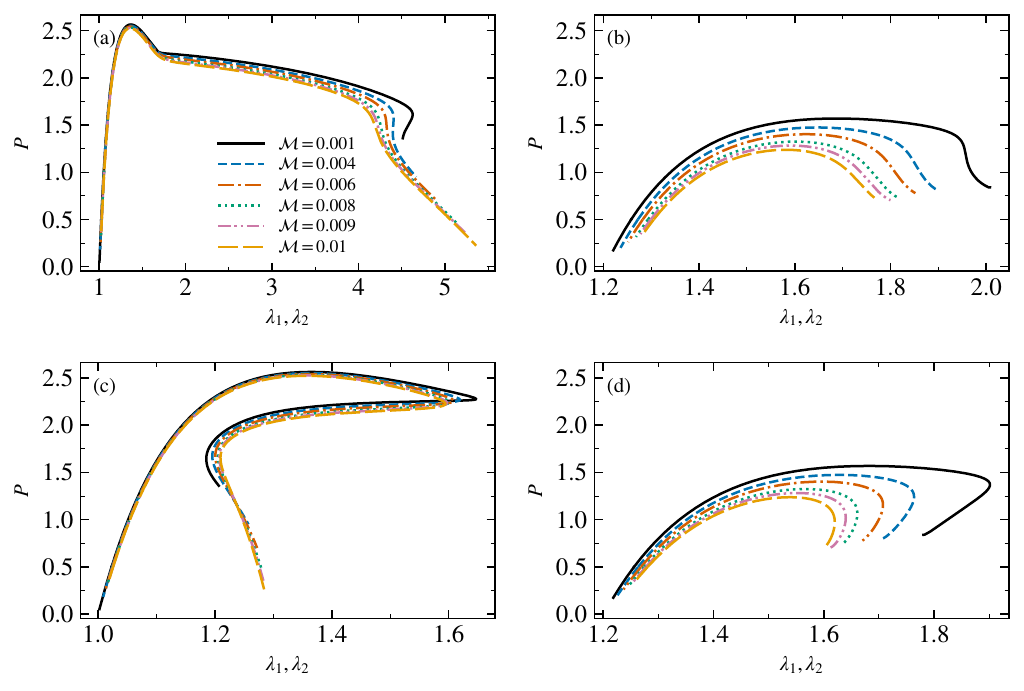}
    \caption{Polar pressure--stretch curves, for unequal radii ($a_\mathrm{u} = \sqrt{3}$, $a_\mathrm{l} = \sqrt{2}$) and various $\mathcal{M}$.}
    \label{fig:P_vs_lambda_poles_TC_DIff_a}
\end{figure}

Finally, both coils carry equal currents, with $\mathcal{M}_\mathrm{u} = \mathcal{M}_\mathrm{l} = \mathcal{M}$, but have different radii: $a_\mathrm{u} = \sqrt{3}$ and $a_\mathrm{l} = \sqrt{2}$. The asymmetry now comes from the geometry of the coils rather than from their currents. The pressure--stretch curves at the poles are shown in Figs.~\ref{fig:P_vs_lambda_poles_TC_DIff_a}, for several values of $\mathcal{M}$.

\begin{figure}[H]
  \centering                         
  \includegraphics[width=\linewidth]{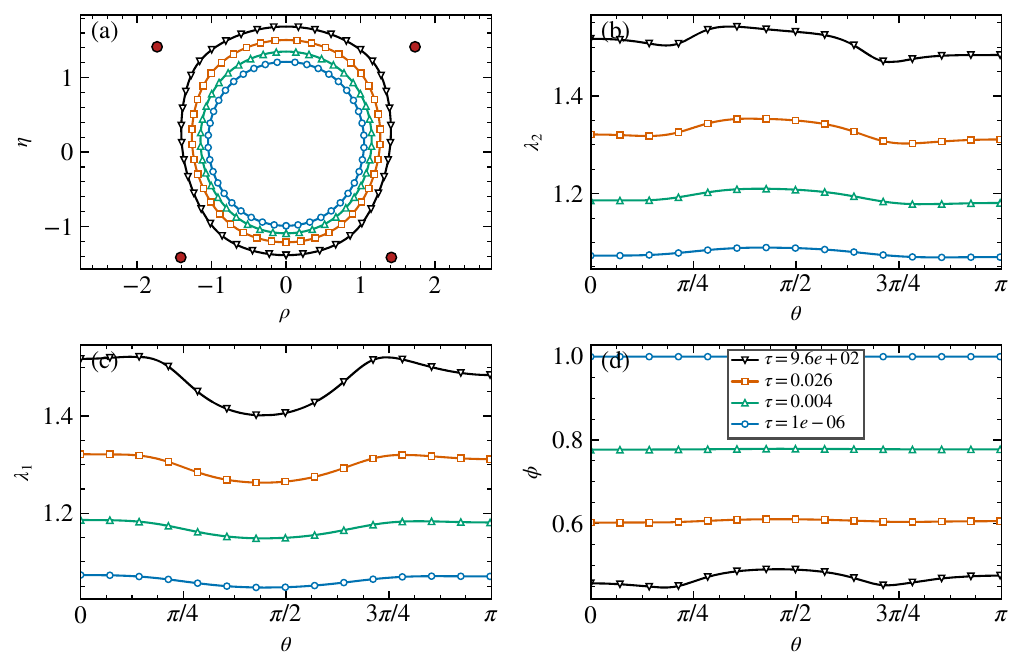}
  \caption{Swelling from the dry state under unequal radii ($a_\mathrm{u} = \sqrt{3}$, $a_\mathrm{l} = \sqrt{2}$, $\mathcal{M} = 0.01$, $P = 1.2$), with the coils at $(\pm a_\mathrm{u}, h)$ and $(\pm a_\mathrm{l}, -h)$; (a)~shape; (b)~circumferential stretch~$\lambda_2(\theta)$; (c)~meridional stretch~$\lambda_1(\theta)$; (d)~polymer volume fraction~$\phi(\theta)$.}
  \label{fig:panel_2x2_m01_TC_diff_a}
\end{figure}

\begin{figure}[H]
    \centering
    \includegraphics[width=\linewidth]{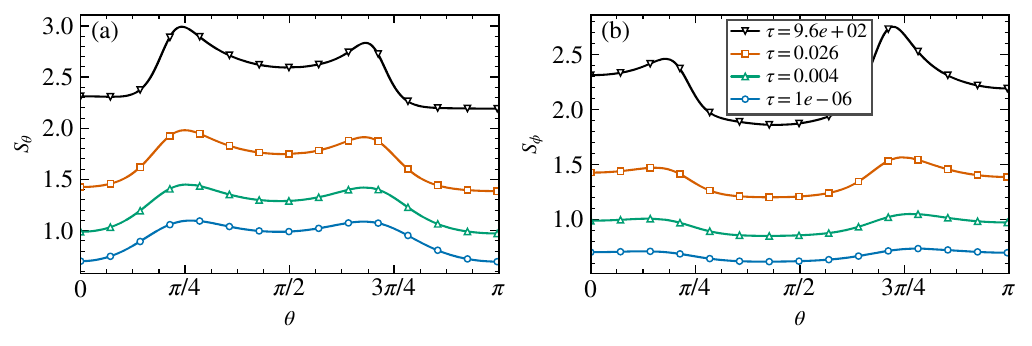}
    \caption{Membrane stresses for the case of Fig.~\ref{fig:panel_2x2_m01_TC_diff_a}: (a)~meridional stress $S_\theta(\theta)$; (b)~circumferential stress $S_\phi(\theta)$.}
    \label{fig:S_panels_m01_TC_diff_a}
\end{figure}

Figs.~\ref{fig:panel_2x2_m01_TC_diff_a} and~\ref{fig:S_panels_m01_TC_diff_a} show the time evolution of swelling at $\mathcal{M} = 0.01$ and $P = 1.2$.

\begin{figure}[H]
    \centering
    \includegraphics[width=\linewidth]{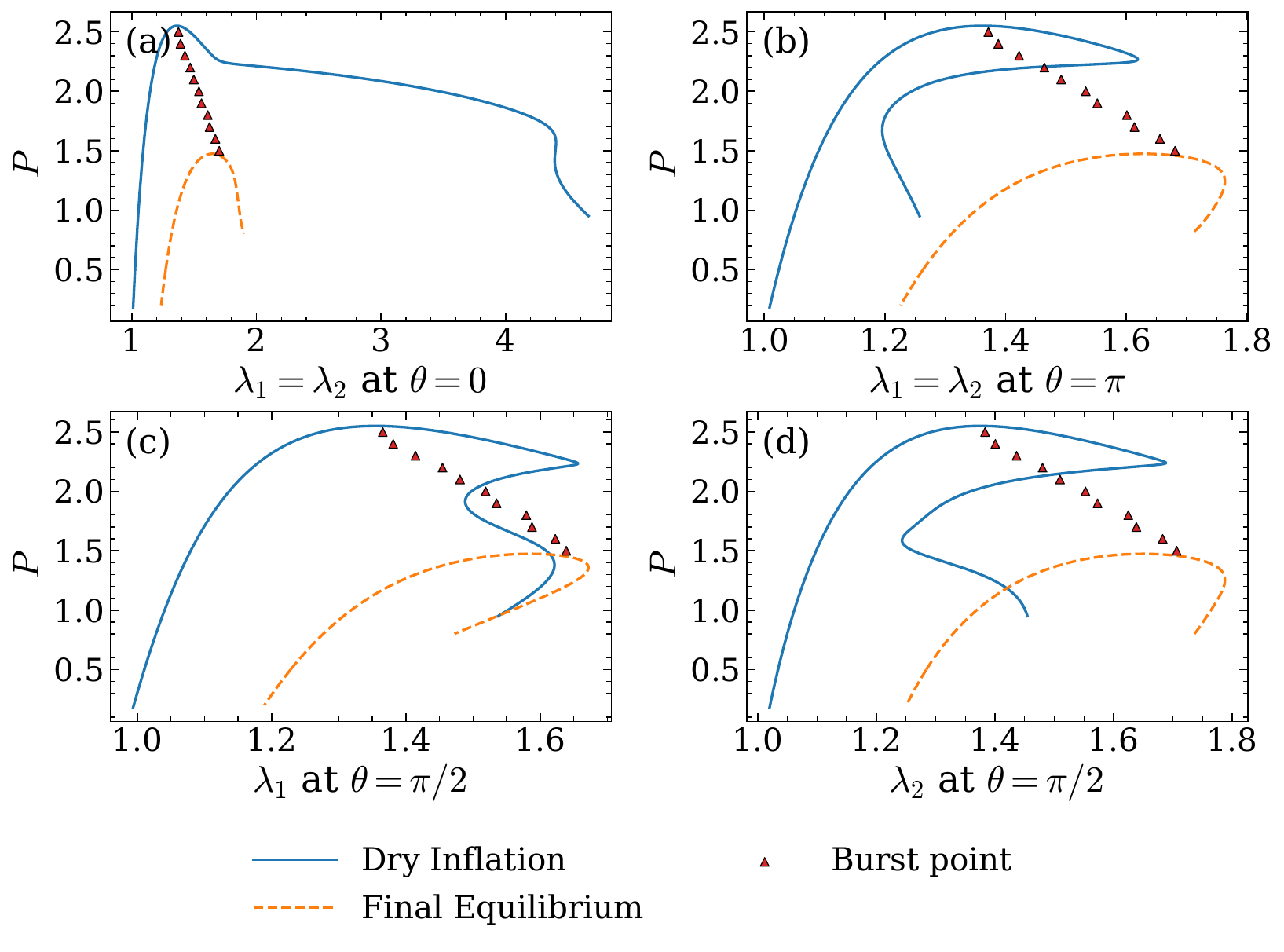}
    \caption{Dry and slow-inflation equilibrium curves with burst points for
    unequal radii ($a_\mathrm{u} = \sqrt{3}$, $a_\mathrm{l} = \sqrt{2}$,
    $\mathcal{M}_\mathrm{u} = \mathcal{M}_\mathrm{l} = 0.004$):
    (a)~$P$ vs.\ $\lambda_1 = \lambda_2$ at $\theta = 0$;
    (b)~$P$ vs.\ $\lambda_1 = \lambda_2$ at $\theta = \pi$;
    (c)~$P$ vs.\ $\lambda_1$ at $\theta = \pi/2$;
    (d)~$P$ vs.\ $\lambda_2$ at $\theta = \pi/2$. Triangles denote burst
    points.}
    \label{fig:Burst_point_a_unequal}
\end{figure}

Figure~\ref{fig:Burst_point_a_unequal} shows the corresponding results for
unequal radii at $\mathcal{M} = 0.004$, with the panels arranged as in
Fig.~\ref{fig:Burst_point_M_unequal}. The dry curve again reaches a maximum
pressure of $P = 2.55$, beyond which the stretch at $\theta = 0$ grows
to $\lambda_1 = \lambda_2 =  4.7$, while the stretch at $\theta = \pi$ remains below
$\lambda_1 = \lambda_2 = 1.6$. On the equator, the dry curve now folds back more than once,
in terms of both $\lambda_1$ and $\lambda_2$. The slow-inflation curve has
its limit point at $P = 1.48$, so the burst points
occurs for pressure spanning $1.5 \leq P \leq 2.5$.

The stretch at burst increases monotonically as the pressure decreases
towards the limit point. At $\theta = 0$, it increases from $\lambda_1 = \lambda_2 = 1.37$
at $P = 2.5$ to $\lambda_1 = \lambda_2 =  1.70$ at $P = 1.5$, and at $\theta = \pi$ from
$\lambda_1 = \lambda_2 =  1.37$ to $\lambda_1 = \lambda_2 =  1.68$. At the equator, $\lambda_1$ increases from
$1.37$ to $1.64$, and $\lambda_2$ from $1.38$ to
$1.71$. The stretches at burst are slightly larger than in the case
of unequal intensities (Fig.~\ref{fig:Burst_point_M_unequal}), but again they 
remain close to each other.

\section{Summary}

This study develops a transient, fully coupled, nonlinear multiphysics finite
element formulation (\textbf{FEM}) for the axisymmetric swelling of
an initially spherical magnetoelastic hydrogel balloon held at constant
pressure and subjected to the non-uniform magnetic field of one or two
current-carrying coils. Large elastic deformation, solvent diffusion through
the crosslinked polymer network, and magnetic loading are resolved
simultaneously within a staggered algorithm using a Taylor--Hood-type
discretisation. Unlike the equilibrium energy-minimisation approach of \cite{TengLi_JMPS2019,TAMADAPU2022},
which resolves only the end states of the process, the present formulation
follows the complete time-dependent evolution of the balloon from its dry
inflated state either to its swollen equilibrium or to burst.

The formulation is verified in three ways. It recovers the analytical
free-swelling stretch, it reproduces the delayed-burst regime reported in
\cite{TengLi_JMPS2019}, and, under magnetic loading, its converged equilibrium
states coincide pointwise with those of shooting method, in shape, in both
principal stretches, and in the polymer volume fraction. This agreement holds
both for a symmetry-preserving initial state ($a = 2$) and for a
symmetry-breaking one ($a = 3$). It confirms that the two formulations share a
common equilibrium limit, and that the additive elastic--mixing--magnetic
decomposition of the free energy used for bulk magnetoelastic gels in
\cite{garcia2020magneto} carries over to the thin-membrane setting.

In the absence of a magnetic field, a dry pear-shaped (symmetry-breaking)
balloon does not remain pear-shaped once exposed to the solvent. It passes
rapidly through a sequence of other pear-shaped configurations and then
relaxes onto the symmetric slow-inflation branch, provided that the applied
pressure lies below the limit-point pressure of the fully swollen balloon. During
this process, the strong stress concentration at the highly stretched pole of
the dry configuration is removed, and the stress becomes uniform over the
membrane. When the starting pressure exceeds the limit-point pressure, the balloon
bursts after a finite time. A balloon starting from a symmetry-breaking
configuration reaches a larger polar stretch at burst than one starting from a
symmetric configuration, but bursts earlier.

Under the field of a single coil placed in the equatorial plane, the balloon
retains its equatorial plane of symmetry throughout swelling, and the
equatorial region is drawn towards the coil. Increasing the field intensity
increases the meridional stretch and the solvent uptake at the equator, while
leaving the poles nearly unaffected, and concentrates the membrane stresses
in the equatorial region, where the magnetic loading is greatest. For a coil
of radius $a = 2$ and $\mathcal{M} \geq 0.006$, the slow-inflation curve
becomes a closed loop, so swollen equilibria exist only within a
bounded window of pressure, which narrows from both above and below as
$\mathcal{M}$ increases. Outside this window the balloon bursts, and the burst
points form two families, above and below the closed loop. For a coil of radius
$a = 3$, a dry symmetry-breaking configuration swells into an equilibrium
state that is symmetric about the equator for all field intensities
considered.

Under two coaxial coils, the balloon deforms locally near each coil. With
identical coils carrying equal currents, the shape, stretches, solvent
fraction, and stresses remain symmetric about the equator throughout swelling.
With unequal currents or unequal radii, the loading itself is asymmetric, and
the response remains asymmetric about the equator throughout swelling, with
the dominant deformation and stress peaks near the more strongly loading coil.
In all the single- and two-coil cases examined with weak fields, the equilibrium curve of fully swollen state is not an isola, burst occurs above its limit-point
pressure, and the stretch at burst increases as the pressure decreases towards
this limit point. Also burst is observed below the magnetic limit-point which is at small pressures and stretches. 

\section{Conclusion}

In this work interesting observation are made from the analysis of magneto-elastic behavior of gel balloon. 
First, in all the cases considered, dry 
configurations that break the symmetry do not persist under swelling. They undergo rapid 
rearrangement followed by diffusion-driven relaxation onto the symmetric
branch. The asymmetry about the equator in the swollen state arises only when the
loading imposes it, through unequal coil currents or radii. Second, the
magnetic field controls whether stable swelling is possible. For coils which are sufficiently
closer to undeformed balloon, beyond a critical field intensity, the slow-inflation curve closes
into a loop. Stable swollen states then exist only within a bounded pressure
window that narrows as the field increases, and outside this window the
balloon bursts. Third, transient membrane stresses can far exceed their
equilibrium values. The formulation also yields the time to burst. In
thin-walled active gel structures, the most severe loading may therefore occur
well before swelling is complete.

These conclusions are bounded by the assumptions of the model. The analysis is
axisymmetric by construction, so non-axisymmetric modes are excluded rather
than ruled out. The solvent diffusivity is taken to be constant and isotropic,
the mobility factor $\gamma(\phi)$ is assumed to be linear in the polymer volume
fraction, and the coil field is prescribed rather than solved self-consistently
with the deformed magnetized body. Future work will relax these assumptions
and extend the framework to thermally responsive gels, to coupled
electro--magneto--mechanical--diffusion problems, and to non--spherical shells
such as ellipsoidal, cylindrical and toroidal geometries. A further direction
is to reimplement the formulation within a JAX-based differentiable finite
element framework. Automatic differentiation would then give gradients of the
transient response with respect to coil currents, coil geometry, pressure and
material parameters. This would enable inverse design of magnetically actuated
gel balloons, for example finding coil configurations that produce a target
swollen shape or keep transient stresses below a failure threshold.

\section{Appendix} 
\section*{Shooting Method}
\label{sec:Appendix}
 A primary field variable $\bar{\rho}$ is replaced with $\lambda_2$ as the new primary field variable by doing the following substitutions, 
\begin{equation}
    \bar{\rho} = \lambda_2 R_0 \sin \theta, \quad \lambda_{2,\theta} \sin\theta = v,  \quad \bar{w} = \bar{\eta}_{,\theta}, \quad \lambda_1 = \frac{ \sqrt{ \left(\lambda_2\cos\theta + v\right)^2R_0^2 + \bar{w}^2 } }{R_0},
    \label{subs1}
\end{equation} 
The total potential energy is non-dimensionalised using $\mu^*$, and the non--dimensional quantities are given by  
\begin{equation}
 \left({\rho}, {\eta} , w, {a} \right)= \frac{\left(\bar{\rho}, \bar{\eta}, \bar{w}, \bar{a} \right)}{R_0}\quad {{S}_\text{e}} = \frac{\bar{S}_{\text{e}}}{\mu^*}\quad {P} = \frac{\bar{P}R_0}{T_0 \mu^*}, \quad \pi = \frac{\bar{\pi}}{\mu^*}, \quad   \mu_i = \frac{\bar{\mu}_i}{\mu^*},  
 \label{nd1} 
\end{equation} 
where $\bar{\mu_i}$ and $\mu_i$ are dimensional and non-dimensionalised Ogden material parameters  given in (\ref{energy_Ogden1}). Using equations \eqref{subs1} and \eqref{nd1} total potential energy can be written as, 

\begin{equation*} 
    \pi = \int_0^{2\pi} \int_0^{\pi} \hat{\Pi} \,   {\rm d}\theta \,{\rm d}\phi,
\end{equation*}
where the total potential energy density $\hat{\Pi}$ is given by

\begin{equation} 
\label{Total Energy nd}
    \hat{\Pi} =   \left( S_{\text{e}} - S_{\text{mix}} -\frac{J}{2}  \chi_m  \mathcal{M} \mathcal{H}^2 \right)\sin\theta + \frac{1}{2}{P}\lambda_2^2{{\eta}}_{,\theta}\sin^2\theta.   
\end{equation}  

The total potential energy functional now becomes $\pi = \pi\left( \lambda_2 , v, \eta, w , J,\theta \right)$. Using the principle of minimum potential energy, equations of equilibrium are obtained as 
\begin{align}
    \label{equilibrium} 
    \frac{\partial \hat{\Pi}}{\partial \lambda_2} - \frac{\rm d}{\rm d \theta} \left( \frac{\partial \hat{\Pi}}{\partial v} \sin\theta \right) =0, \qquad  \frac{\partial \hat{\Pi}}{\partial \eta} -  \frac{\rm d }{\rm d \theta} \frac{\partial \hat{\Pi}}{\partial w} = 0.  \qquad  \frac{d}{d\theta}\frac{\partial \hat{\Pi}}{\partial J} = 0 .  
\end{align}
The boundary conditions for the axisymmetric inflation of a balloon are given as
\begin{align} \label{BC1}
    \rho\left( 0 \right) = 0, \quad \rho\left( \pi \right) = 0, \quad \eta_{,\theta} \left( 0 \right) = 0, \quad \eta_{,\theta} \left( \pi \right) = 0.
\end{align} 
Substituting \eqref{subs1} in \eqref{equilibrium} converts a set of three second-order ordinary differential equations \eqref{equilibrium} to a set of five first-order ordinary differential equations given as 
\begin{align}
   & \lambda_{2,\theta} =  \frac{v}{\sin\theta},\label{gdef1}\\   
   & v_{,\theta} = F_1\left( v, \eta, w, \lambda_2 , J,\theta; P \right),\label{gdef2}\\
    & w_{,\theta} = F_2\left( v,\eta,w , \lambda_2 , J, \theta; P \right), \label{gdef3}\\ 
    & J_{,\theta} = F_3\left( v,\eta,w , \lambda_2 , J, \theta; P \right),  \label{gdef4} \\ 
    & \eta_{,\theta} = w \label{gdef5}, 
\end{align} 
with the boundary conditions in terms of new field variables as,  
\begin{equation}\label{bc}
 v(0)=0,\quad w(0)=0,\quad v(\pi)=0,\quad w(\pi)=0.
\end{equation} 
Due to the coordinate singularity at the poles ($\theta=0$ and $\theta=\pi$), the right hand side of the equations \eqref{gdef1}-\eqref{gdef4} attain an indeterminate $0/0$ form and are thus valid only for $\theta\neq 0 $ and $\theta \neq \pi$. Applying L'Hôpital's rule to indeterminate form,  equations set valid for the poles are given as, $\lambda_{2,\theta} = 0 $, $v_{,\theta} = 0$, $w_{,\theta} = \left( - J\chi_m  \mathcal{M} \mathcal{H} \mathcal{H}_{,\eta} -P\lambda_2^2\cos\theta  \right)/\left(2{S_{\text{e}}}_{,ww} \right) $ , $J_{,\theta} = 0 $ and $\eta_{,\theta}=0$. (see     for details) 

\subsection{Solution procedure} 
Following the approach outlined in recent work \citep{shaikh2025inflation,roychowdhury2015inflating, roychowdhury2015response, TAMADAPU_IJNM2022, TAMADAPU_Euro_A, TAMADAPU_IJSS_rim}, the two-point boundary value problem (TPBVP) defined by \eqref{gdef1}--\eqref{bc} for the axisymmetric inflation of a magnetoelastic balloon is solved using a shooting method, which recasts the TPBVP as an initial value problem (IVP). However, due to the removable singularity at the poles, direct numerical integration from $\theta = 0$ to $\pi$ is not feasible. To circumvent this, the equations are integrated separately from $\theta = 0$ to $\pi/2$ and from $\theta = \pi$ to $\pi/2$, hereafter referred to as the \emph{forward} and \emph{backward} integrations of \eqref{gdef1}--\eqref{bc}, respectively.

For a given $(P, \mathcal{M})$ pair, the forward integration requires $\lambda_2(0)$, $\eta(0)$, and $J(0)$, while the backward integration requires $\lambda_2(\pi)$, $\eta(\pi)$, and $J(\pi)$. The values of $J(0)$ and $J(\pi)$ are obtained from the fact that the principal stretches coincide at the poles, $\lambda_1 = \lambda_2 = \lambda_{\text{pole}}$, together with the plane-stress condition for the membrane, $\partial \hat{\Pi}/\partial J = 0$ (which physically expresses the vanishing of the stress, $\sigma_{rr} = 0$). For a chosen value of $\mathcal{M}$, this condition reduces to a relation involving only $\lambda_{\text{pole}}$ and $J$, from which $J(0)$ and $J(\pi)$ can be computed once $\lambda_2(0)$ and $\lambda_2(\pi)$ are specified. Accurate values of $\lambda_2(0)$, $\eta(0)$,  $\lambda_2(\pi)$ and $\eta(\pi)$ must therefore satisfy the matching conditions at $\theta = \pi/2$:
\begin{align*}
    &\lambda_2^f(\pi/2) = \lambda_2^b(\pi/2), \quad v^f(\pi/2) = v^b(\pi/2), \quad w^f(\pi/2) = w^b(\pi/2), \\
    &\eta^f(\pi/2) = \eta^b(\pi/2), \quad J^f(\pi/2) = J^b(\pi/2),
\end{align*}
where the superscripts $f$ and $b$ denote the forward ($\theta = 0 \to \pi/2$) and backward ($\theta = \pi \to \pi/2$) integrations, respectively. This matching requirement is cast as an optimisation problem parameterised by the pair $(\Delta, \Gamma) = (\lambda_2(0), P)$, with optimisation variables $\bigl(\Delta, \eta(0),  \lambda_2(\pi), \eta(\pi)\bigr)$ and cost function
\begin{align}
\label{e:cost}
S&\left(\Delta, \eta(0), \lambda_2(\pi), \eta(\pi); \Gamma\right) = \Bigl[\{\lambda_2^f(\pi/2) - \lambda_2^b(\pi/2)\}^2 + \{v^f(\pi/2) - v^b(\pi/2)\}^2 \Bigr. \nonumber \\
&\Bigl. \quad + \{w^f(\pi/2) - w^b(\pi/2)\}^2 + \{\eta^f(\pi/2) - \eta^b(\pi/2)\}^2 + \{J^f(\pi/2) - J^b(\pi/2)\}^2 \Bigr]^{1/2}.
\end{align}
Here, $\Gamma$ acts as the numerical control parameter while all other system parameters are held fixed. The choice $(\Delta, \Gamma) = (\lambda_2(0), P)$ is convenient for resolving portions of the pressure–stretch curve along which the pressure varies rapidly with a small change in stretch. Conversely, the alternative choice $(\Delta, \Gamma) = (P, \lambda_2(0))$ in \eqref{e:cost} is better suited to portions along which the stretch varies rapidly with a small change in pressure. For each value of $\Gamma$, the cost function \eqref{e:cost} is minimised until $S$ falls below a prescribed tolerance, and the converged solution is used as the initial guess for the next increment of $\Gamma$. Repeating this procedure traces the complete equilibrium branch, whose converged states are compared with those of the transient finite element formulation in the main text.

\bibliographystyle{abbrvnatX2}    
\bibliography{cas-refsX-2}
\end{document}